\documentclass[preprint,aip,pop]{revtex4-1}
\usepackage{amsmath}
\usepackage{epsfig,wrapfig}
\usepackage{subfig}
\usepackage{amsfonts}
\begin{document}

\begin{abstract}
Ion sound instabilities driven by the ion flow in a system of a finite
length are considered by analytical and numerical methods. The ion sound
waves are modified by the presence of stationary ion flow resulting in
negative and positive energy modes. The instability develops due to coupling
of negative and positive energy modes mediated by reflections from the
boundary. It is shown that the wave dispersion due to deviation from
quasineutrality is crucial for the stability. In
finite length system, the dispersion is characterized by the length of the
system measured in units of the Debye length. The instability is studied
analytically and the results are compared with direct, initial value
numerical simulations.
\end{abstract}

\title{Ion sound instability driven by the ion flows.}
\author{O. Koshkarov}
\email{koshkarov.alexandr@usask.ca}
\affiliation{ Department of Physics and Engineering Physics, University of Saskatchewan \\ 116 Science Place, Saskatoon, SK S7N 5E2, Canada} 
\author{A. I. Smolyakov}
\affiliation{ Department of Physics and Engineering Physics, University of Saskatchewan \\ 116 Science Place, Saskatoon, SK S7N 5E2, Canada}
\affiliation{National Research Centre (NRC "Kurchatov Institute"), Moscow, Russia}
\author{I. D. Kaganovich}
\affiliation{Princeton Plasma Physics Laboratory, Princeton, New Jersey, 08543, USA }
\author{V. I. Ilgisonis}
\affiliation{National Research Centre (NRC "Kurchatov Institute"), Moscow, Russia}
\maketitle
\section{Introduction}

Many natural settings of space and laboratory plasmas often include
equilibrium flows of ions and/or electrons. Such situations occurs in
various plasma devices for electric propulsion, plasma diodes, plasma
accelerators, plasma processing devices, and emissive probe diagnostics.
Plasmas permeated by energetic beams is also typical situations in space and
astrophysics \cite{Vranjes}. Such plasmas represent a typical example of a
non-equilibrium system prone to instabilities due to presence of free energy
reservoir from stationary flows. One of the simplest examples is the
excitation of ion-sound waves when the relative velocity between electrons
and ions exceeds the ion sound velocity, $v_{0}>c_{s}.$ \cite%
{VTP7,Skiff,IonInst}. In infinite plasma, the instability may occur as a
result of the kinetic interaction of electrons with the ion beam (two-stream instability due to inverse
Landau damping). On other hand, a number of practical plasma configurations
have the finite length and it is of interest to investigate the
modification/new regimes of instabilities related to the presence of
boundaries. Instabilities due to accelerated ion flows are of interest for
the sheath region of the plasma-material boundaries \cite{FAROUKI}, plasma
diodes \cite{diod_rev,SCHAMEL_eq}, double layers\cite{Keesee,Baalrud,DL1,DL2}%
, and electric propulsion systems \cite{KapulkinThrusters}. In an infinite
plasma, the stationary ion flow $v_{0}$ results in the Doppler shift of the ion sound waves frequency, $\omega \rightarrow \omega
-kv_{0}.$ It is shown in our paper that in a finite length systems, the ion
sound waves can be destabilized due to reflections from the boundaries and
coupling with ballistic modes, $\omega =kv_{0},$ supported
by the ion flow. This instability is different from the above noted
two-stream type ion sound instability where the kinetic resonances is
important.

The Pierce plasma diode \cite{Pierce} is a well studied case of the
instability driven by electron flow in a a finite length system. Various
extensions of the instabilities in Pierce-like plasma systems and related
numerical and experimental studies have been discussed in the literature 
\cite{diod_rev,apl_Cary,apl_Kuhn,apl_Crystal,apl_Godfrey,Matsumoto}. It is
shown in our paper that the problem of the ion sound waves in a system with
boundaries, in a special limit of strong dispersion, is formally reduced to
the Pierce like equations.

In our model we consider only fluid (hydrodynamic) effects, ions are assumed
to be cold (to avoid Landau damping) and have an uniform velocity with
respect to the electron component. Electrons are assumed to be in Boltzmann
equilibrium (electron inertia effects are neglected). We employ analytical
and numerical methods to analyze the structure of unstable eigen-modes,
determine the dispersion relations and conditions for the instability, and
find the frequencies and growth rates of the unstable modes.

\section{Overview of basic equations and instability mechanism}

In this section we present basic equations describing the ion sound waves in
a finite length system and give an overview of the instability mechanism.
The dynamics of cold ions is described by linearized hydrodynamic equations 
\begin{subequations}
\label{full_system}
\begin{align}
& \frac{\partial n_{i}}{\partial t}+v_{0}\dfrac{\partial n_{i}}{\partial z}%
+n_{0}\dfrac{\partial v_{i}}{\partial z}=0, \\
& \frac{\partial v_{i}}{\partial t}+v_{0}\dfrac{\partial v_{i}}{\partial z}+%
\dfrac{e}{m_{i}}\dfrac{\partial \phi }{\partial z}=0.
\end{align}%
The electrons are assumed to be adiabatic and follow Boltzmann relation
assuming low frequency fluctuations, $\omega <kv_{Te}$, 
\end{subequations}
\begin{equation}
n_{e}=\dfrac{n_{0}e}{T_{e}}\phi .  \label{ne}
\end{equation}%
The system is closed by the Poisson equation 
\begin{equation}
\dfrac{\partial ^{2}\phi }{\partial z^{2}}=-4\pi e\left( n_{i}-n_{e}\right) ,
\label{poisson}
\end{equation}%
where $n_{i}$,$n_{e}$,$\phi $ are the perturbations of the ion, the electron
density and the electrostatic potential respectively, $n_{0}$ - equilibrium
density, $e$,$m_{i}$ - charge and mass of ions, $T_{e}$ - electron
temperature, $v_{0}$ - speed of ion flow, $v_{Te}^{2}=2T_{e}/m_{e}$ -
electron thermal velocity.

For the ion injection from the left boundary, the boundary conditions similar
to the Pierce problem\cite{Pierce} are used%
\begin{equation}
\phi (z=0)=\phi (z=L)=n_{i}(z=0)=v_{i}(z=0)=0,  \label{boundary_conditions}
\end{equation}%
where $L$ - length of the system. The important feature of these boundary
conditions is absence of density and velocity perturbations from the
emitting boundary, e.g. as in double layer devices \cite{DL2} where
accelerated ions are extracted from from the plasma source chamber.

\begin{figure}[th]
\subfloat[$time = 0$]{\includegraphics[width = 80mm]{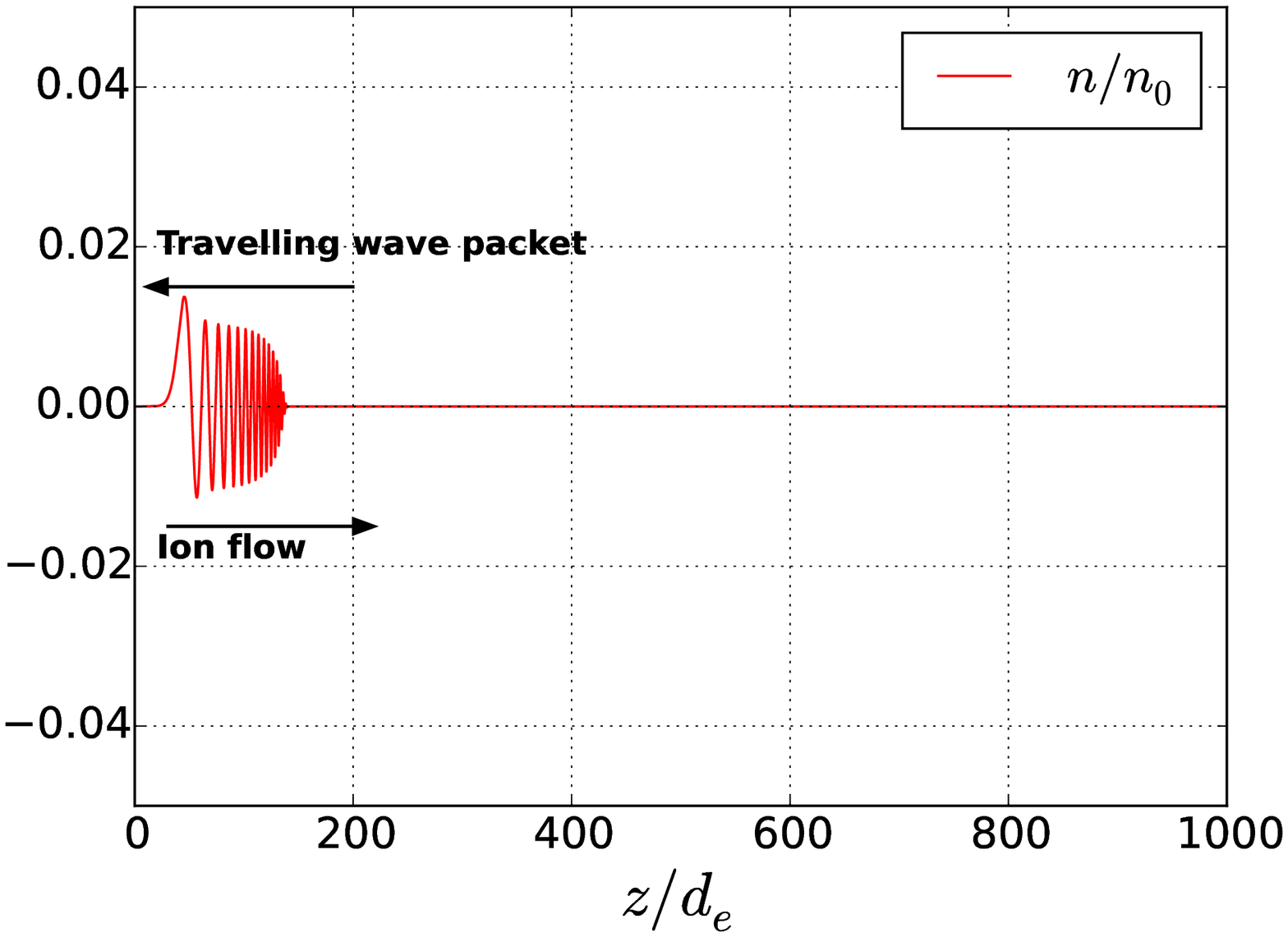}} 
\subfloat[$time = 320\omega_{pi}^{-1}$]{\includegraphics[width =
80mm]{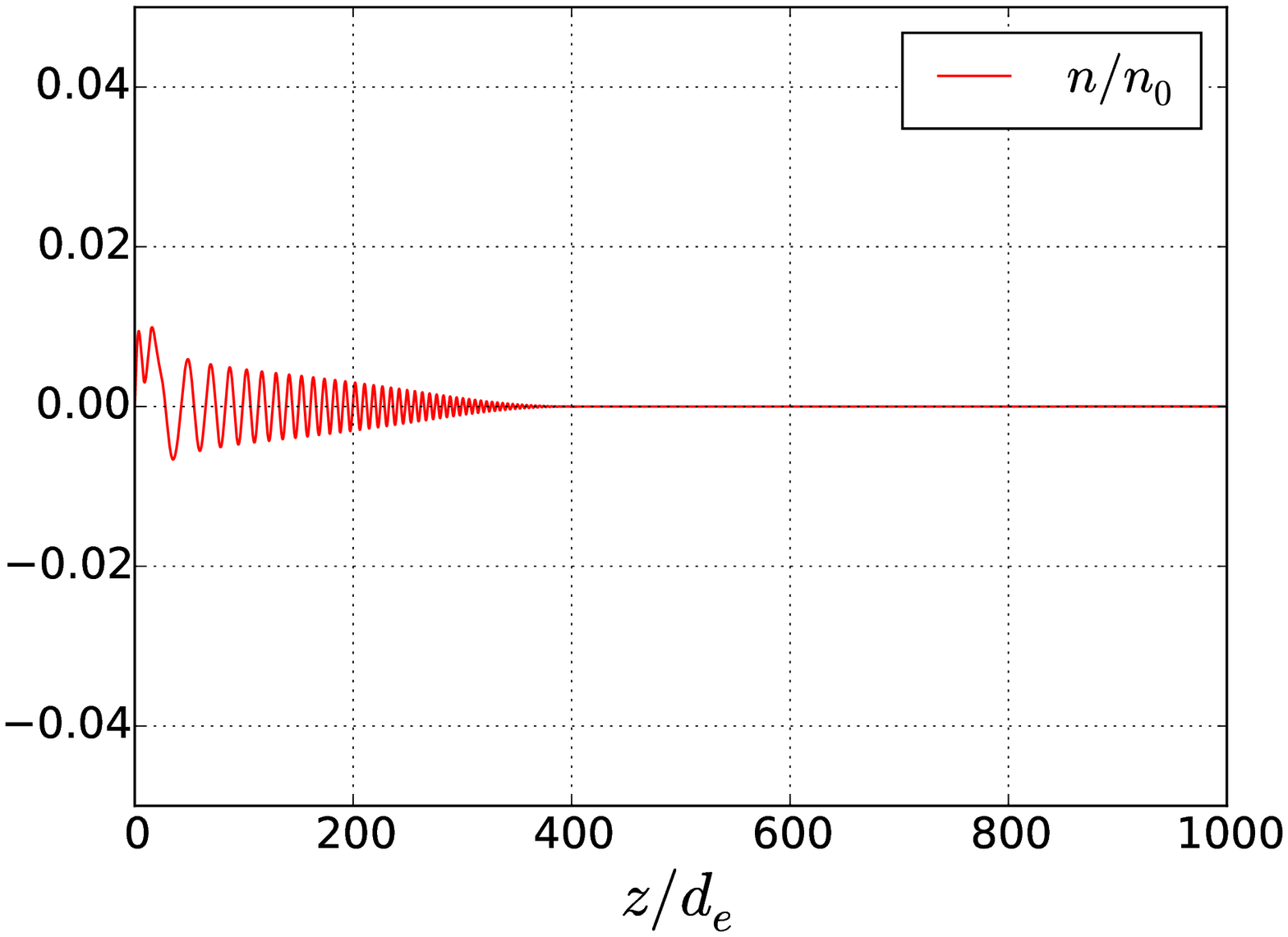}} \newline
\subfloat[$time = 700\omega_{pi}^{-1}$]{\includegraphics[width =
80mm]{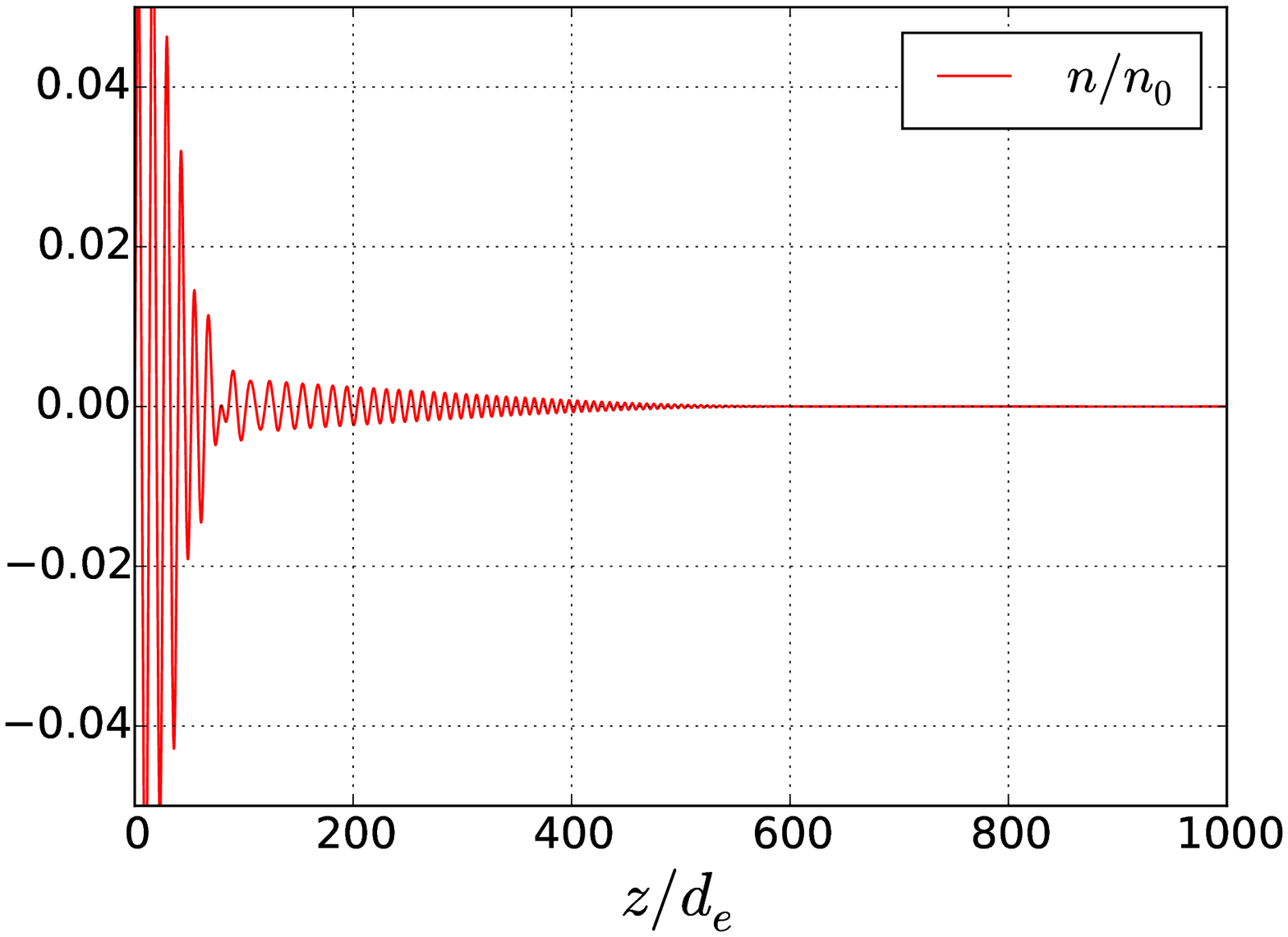}} 
\subfloat[$time =
1000\omega_{pi}^{-1}$]{\includegraphics[width = 80mm]{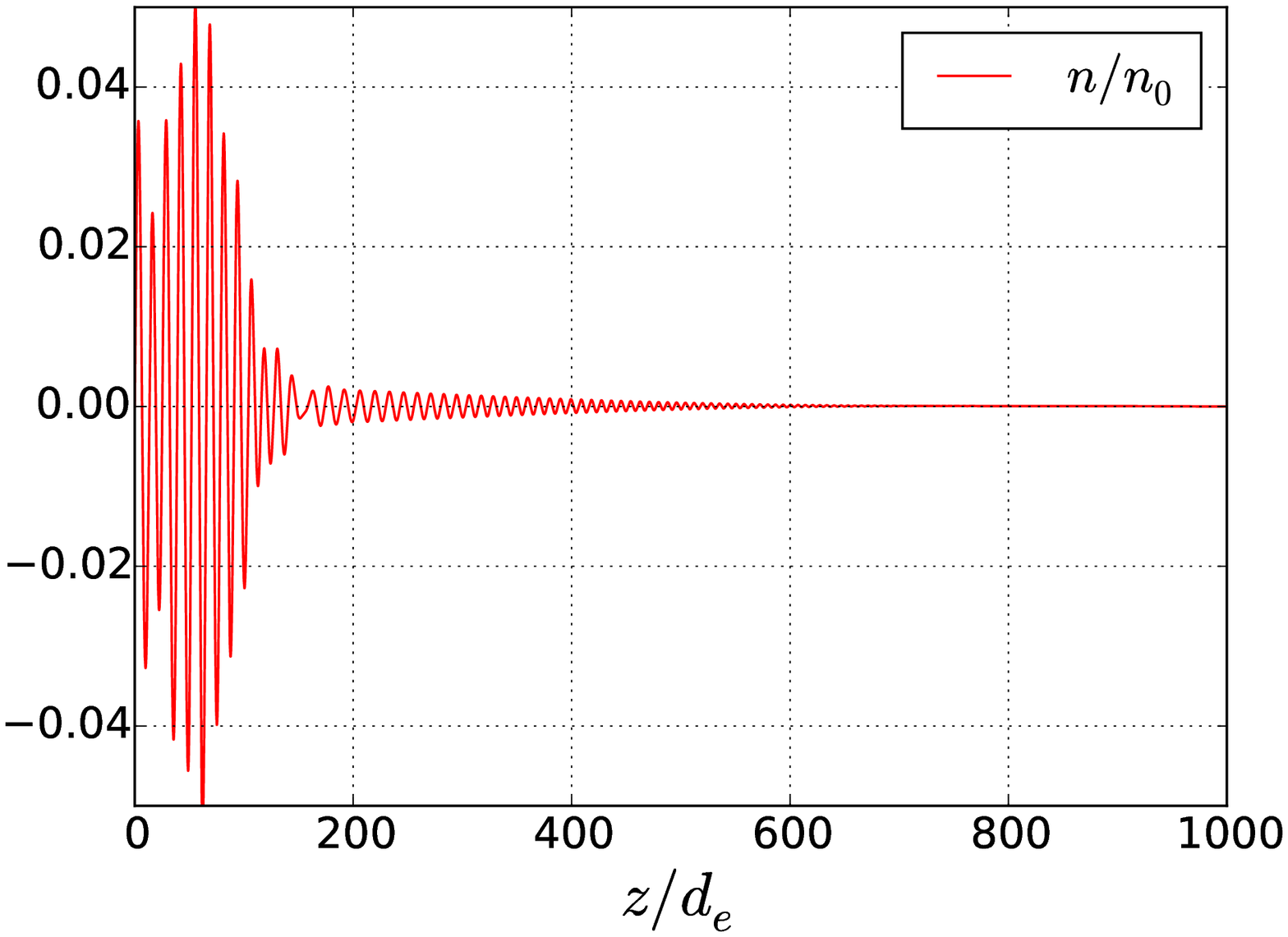}}
\caption{Formation of unstable eigenfunction due to reflection of the wave
packet from the emitting boundary on the left. }
\label{fig:inst_mech:left}
\end{figure}

Ion sound waves on the background of the equilibrium ion flow are described
equations (\ref{full_system}), (\ref{ne}) and (\ref{poisson}). For infinite
length system (periodic boundary conditions), Eqs. (\ref{full_system}), (\ref%
{ne}) and (\ref{poisson}) result in the permittivity 
\begin{equation}
\varepsilon (\omega ,k)=1+\frac{1}{k^{2}d_{e}^{2}}-\frac{\omega _{pi}^{2}}{%
(\omega -kv_{0})^{2}},  \label{eps}
\end{equation}%
where $d_{e}^{2}=T_{e}/(4\pi e^{2}n_{0})$ is the Debye length and $\omega
_{pi}^{2}=4\pi e^{2}n_{0}/m_{i}$ is the ion plasma frequency, $\omega $,$k$
are the frequency and wave number, respectively. The wave mode energy
corresponding to (\ref{eps}) is 
\begin{equation}
\mathcal{E}(\omega ,k)=\omega \frac{\partial \varepsilon }{\partial \omega }%
|k\phi |^{2}=\frac{2k^{2}\phi ^{2}\omega \omega _{pi}^{2}}{(\omega
-kv_{0})^{3}},
\end{equation}%
It follows that the Doppler shift due to the ion flow results in negative
energy perturbations for $\omega <kv_{0}.$ Coupling of negative and positive
energy modes results in reactive instabilities \cite{lashmoreNW,Nezlin}.
In our case, the mode coupling occurs due to boundary conditions on the left
wall as illustrated in Fig. \ref{fig:inst_mech:left}. In Fig. \ref%
{fig:inst_mech:left}a and \ref{fig:inst_mech:left}b traveling wave packet
arrives at the left boundary and starts forming the reflected wave. Further
interaction of the reflected and original waves forms an unstable mode with
an increasing (in time) amplitude as is shown in Figs. \ref%
{fig:inst_mech:left}c and \ref{fig:inst_mech:left}d.

\begin{figure}[th]
\subfloat[$time = 0$]{\includegraphics[width = 54mm]{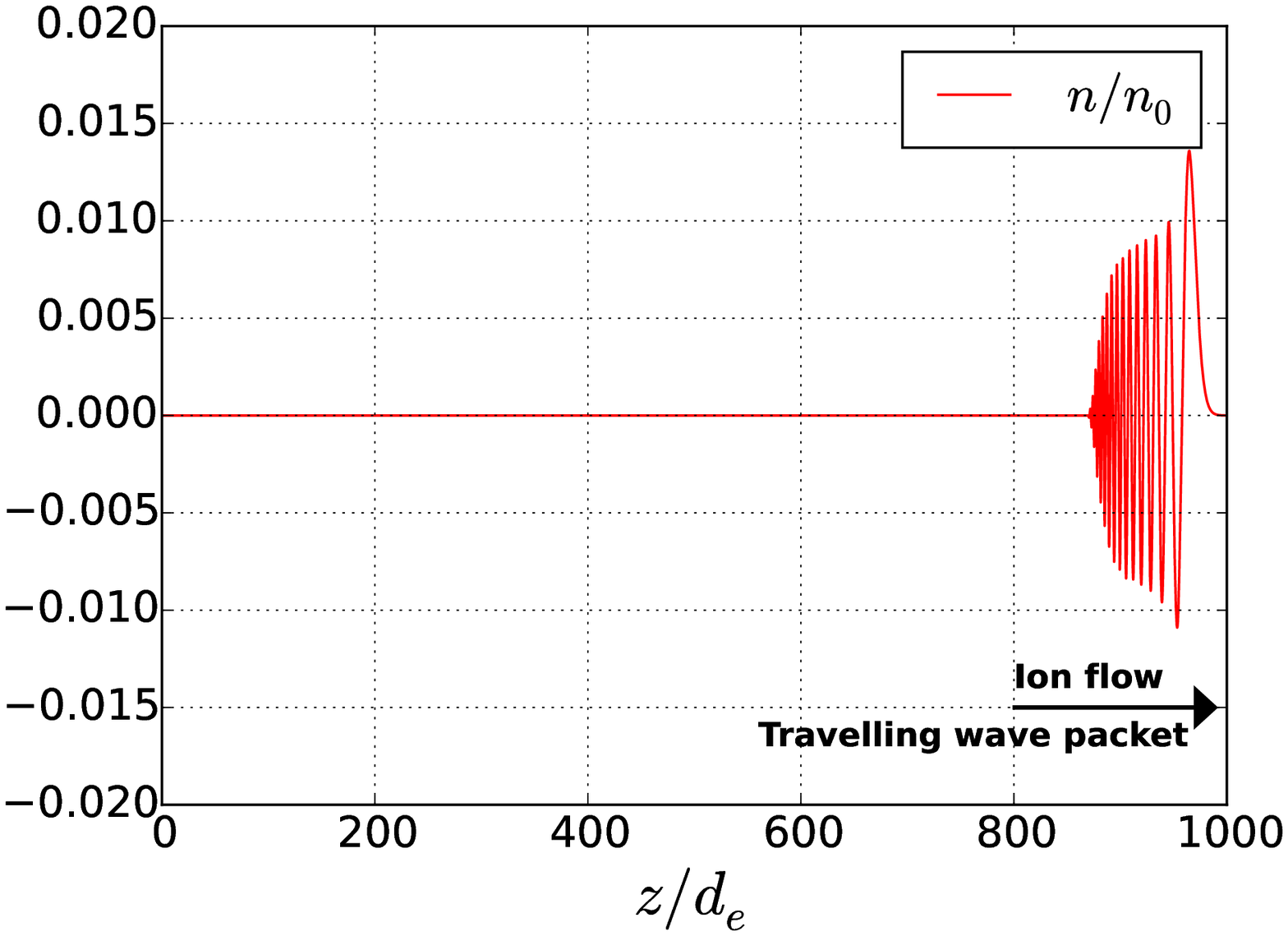}} 
\subfloat[$time = 70\omega_{pi}^{-1}$]{\includegraphics[width =
54mm]{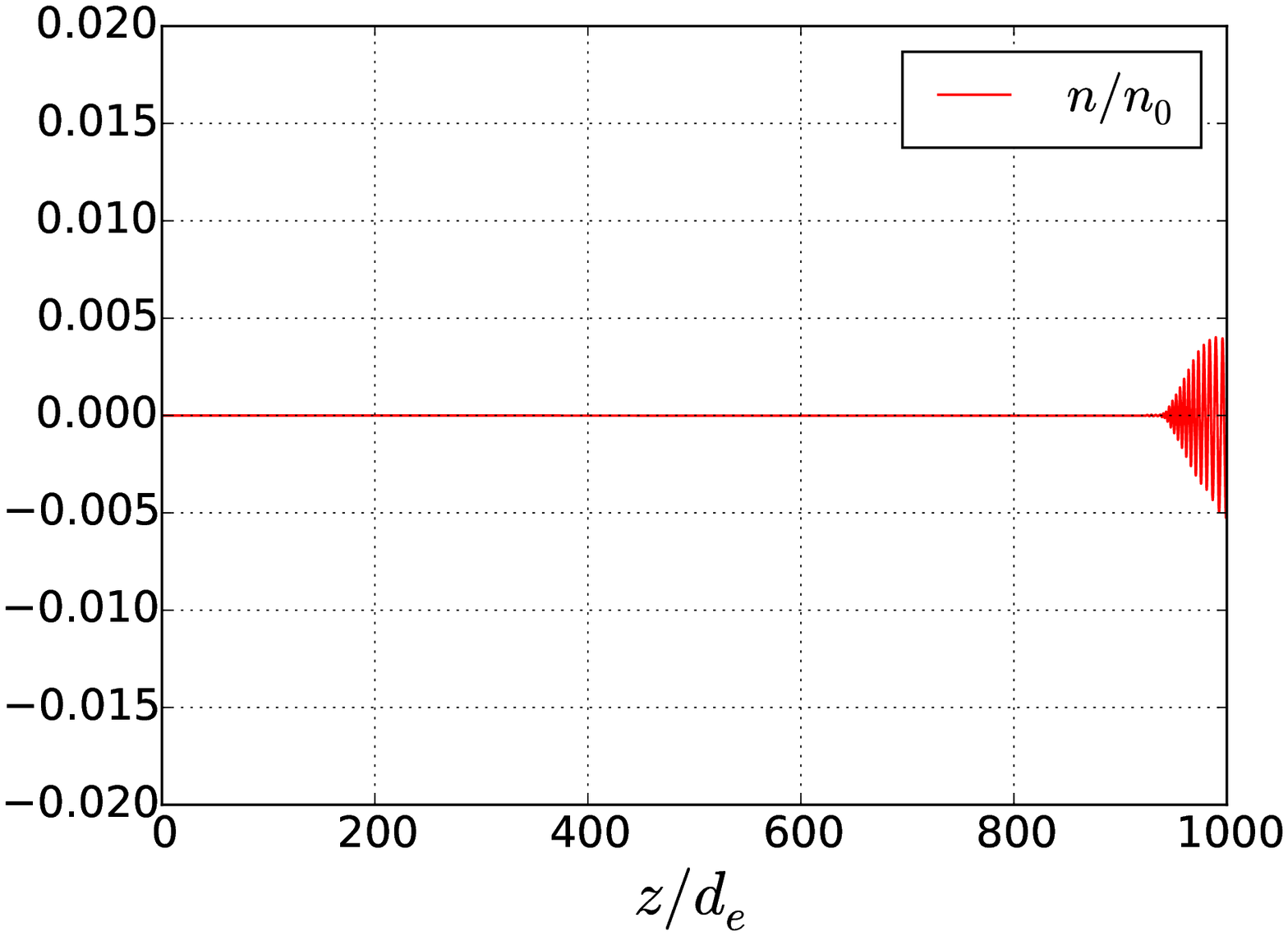}} 
\subfloat[$time =
140\omega_{pi}^{-1}$]{\includegraphics[width = 54mm]{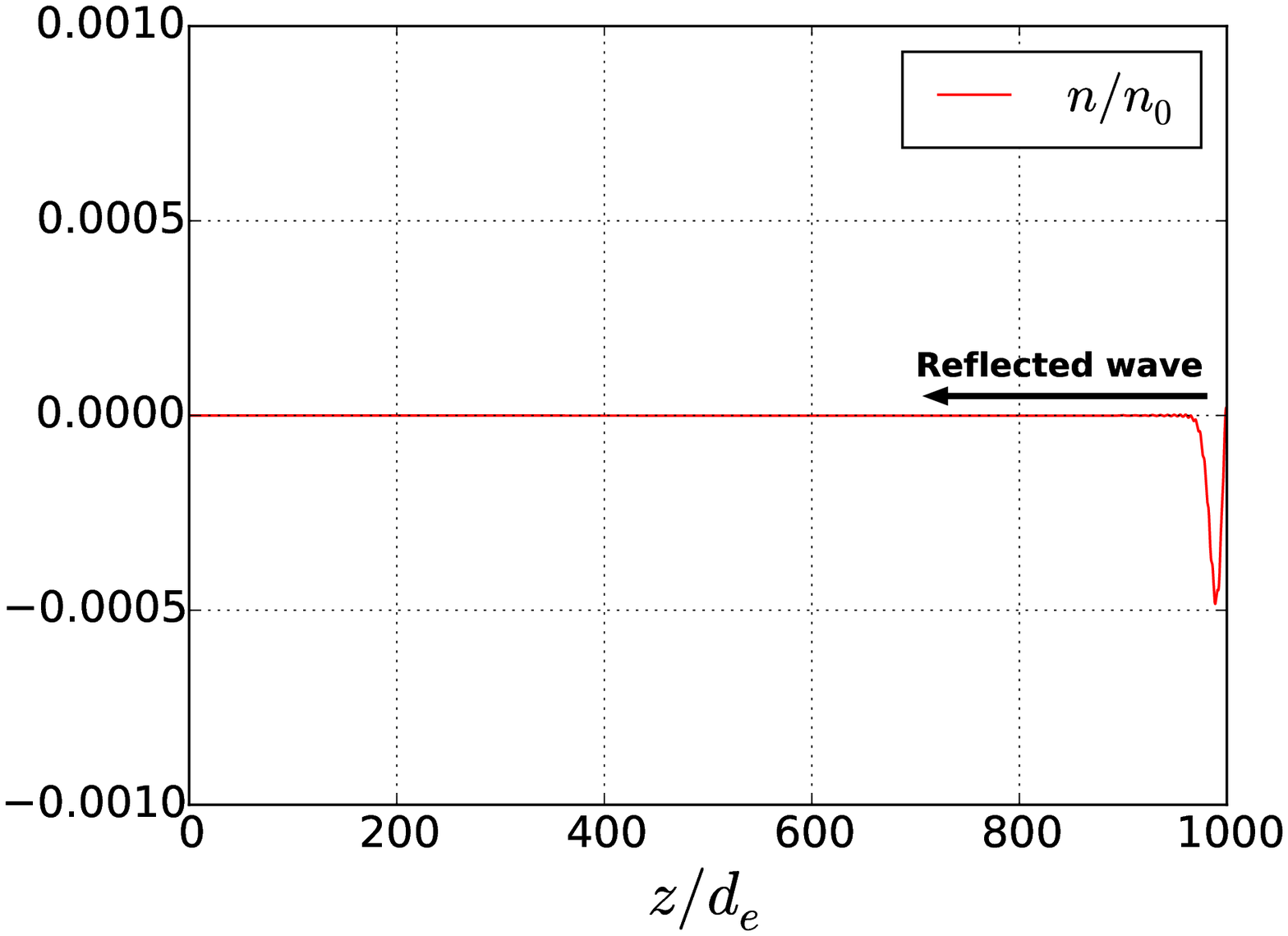}}
\caption{Reflection from the boundary with free density and velocity
perturbations (on the right). }
\label{fig:inst_mech:right}
\end{figure}

The right boundary (with impinging ion flow), where only the potential is
fixed, produce very little reflection, so that the reflected wave amplitude
is much smaller than that of the incident wave (note the different scale in
Fig.\ref{fig:inst_mech:right}c). There is no instability for the reflection
from such a boundary as is illustrated in Fig. \ref{fig:inst_mech:right}.

In next sections we consider the analytical solution for the unstable
eigen-modes, investigate the instability conditions, describe the numerical
method and compare the analytical and numerical results.

\section{Analytical solutions for the eigen-modes}

To study the linearized system (\ref{full_system}) analytically we seek the
solution in the form $\sim e^{-i\omega t}$. Then, the equations (\ref%
{full_system}-\ref{poisson}) can be reduced to a single equation in the
form 
\begin{equation}
v_{0}^{2}\phi ^{\prime \prime \prime \prime }-2i\omega v_{0}\phi ^{\prime
\prime \prime }+\left[ \frac{c_{s}^{2}}{d_{e}^{2}}-\omega ^{2}-\dfrac{%
v_{0}^{2}}{d_{e}^{2}}\right] \phi ^{\prime \prime }+\dfrac{2i\omega v_{0}}{%
d_{e}^{2}}\phi ^{\prime }+\dfrac{\omega ^{2}}{d_{e}^{2}}\phi =0,
\label{phi_full}
\end{equation}%
where prime is a derivative with respect to $z$, $c_{s}^{2}=T_{e}/m_{i}$ -
the ion acoustic velocity. In the limit $v_{0}\rightarrow 0,$ for
perturbations of the form $\sim e^{ikz},$ one obtains the dispersion
equation for the standard ion acoustic waves 
\begin{equation}
\omega ^{2}=\frac{k^{2}c_{s}^{2}}{1+k^{2}d_{e}^{2}}.  \label{no_flow_de}
\end{equation}

General solution of (\ref{phi_full}) can be sought as a sum of the
elementary solutions $\phi \sim C_{i}e^{\lambda _{i}z}$ which are subject to
the boundary conditions (\ref{boundary_conditions}). The characteristic
equation for $\lambda $ has the form%
\begin{equation*}
v_{0}^{2}\lambda ^{4}-2i\omega v_{0}\lambda ^{3}+\lambda ^{2}\left[ \frac{%
c_{s}^{2}}{d_{e}^{2}}-\omega ^{2}-\dfrac{v_{0}^{2}}{d_{e}^{2}}\right] +\frac{%
2i\omega v_{0}}{d_{e}^{2}}\lambda +\frac{\omega ^{2}}{d_{e}^{2}}=0,
\end{equation*}%
or in more convenient form 
\begin{equation}
d_{e}^{2}\left( \lambda -\dfrac{i\omega }{v_{0}}\right) ^{2}\left( \lambda
^{2}-\dfrac{1}{d_{e}^{2}}\right) +\dfrac{c_{s}^{2}}{v_{0}^{2}}\lambda ^{2}=0,
\label{eq_lambda}
\end{equation}
which correspond to the equation (\ref{no_flow_de}) with Doppler shift.

\subsection{Full quasineutrality case}

The dispersion plays important role in instability mechanism. For the length
scales much longer than the Debye length, the charge separation is not
important and one can consider the fully quasineutral case, $n_{i}=n_{e},$
corresponding to the absence of the dispersion.

In this limit, the solution of system (\ref{full_system}) can be obtained in
the form
\begin{equation}
\phi (z)=C_{1}\exp \left( \dfrac{i\omega z}{v_{0}+c_{s}}\right) +C_{2}\exp
\left( \dfrac{i\omega z}{v_{0}-c_{s}}\right) .  \label{qn_phi}
\end{equation}%
By imposing boundary conditions (\ref{boundary_conditions}), we obtain the
stable eigen-modes with the real frequencies $\omega$ 
\begin{equation}
\omega _{n}=\pi n\dfrac{v_{0}^{2}-c_{s}^{2}}{Lc_{s}},n\in \mathbb{Z}.
\label{qn_disp}
\end{equation}

\ It is worth to note, that the formula (\ref{qn_disp}) is not valid in the
zero electron temperature limit, $T_{e}\rightarrow 0$, $c_{s}\rightarrow 0$,
because in this case the solution for electrostatic potential will be
different from (\ref{qn_phi})  
\begin{equation}
\phi (z)=(C_{1}+C_{2}z)e^{\frac{i\omega }{v_{0}}z},
\end{equation}%
while boundary conditions will give us the frequency
\begin{equation}
\omega _{n}=\frac{2\pi n}{L}v_{0},n\in \mathbb{Z}.
\end{equation}

Therefore, the non-dispersive waves are stable.\ As it will be shown below,
the wave dispersion is crucial for the instability mechanism.

\subsection{Weak dispersion case}

In the long systems $d_{e}\ll L$, the dispersion is weak $kd_{e}\ll 1$,
where the wave number $k\sim 1/L$. Using dispersion equation for plasma
without flows (\ref{no_flow_de}) one gets the estimates for the mode
frequency
\begin{equation}
\omega \sim kc_{s}\mbox{ or }\omega \sim \frac{d_{e}}{L}\omega _{pi}.
\label{no_flow_aprox}
\end{equation}%
We solve (\ref{eq_lambda}) treating Debye length as a small
parameter, thus it has four roots where two of them are small $\sim O(1)$
and two of them are large $\sim O(1/d_{e})$. The first pair coincide with those
in quasi neutral case
\begin{equation}
\lambda _{1,2}=\frac{i\omega }{v_{0}\pm c_{s}}+O(d_{e}^{2})\sim O(1).
\label{fc_roots_wd}
\end{equation}%
The second pair is
\begin{equation}
\lambda _{3,4}=\pm \frac{i}{v_{0}d_{e}}\sqrt{c_{s}^{2}-v_{0}^{2}}+\frac{%
i\omega c_{s}^{2}}{v_{0}}\frac{1}{c_{s}^{2}-v_{0}^{2}}+O(d_{e})\sim O(\frac{1%
}{d_{e}}).  \label{sc_roots_wd}
\end{equation}%
Since all roots are different we can write the general solution of (\ref%
{phi_full}) in this form
\begin{equation}
\phi (z)=C_{1}e^{\lambda _{1}z}+C_{2}e^{\lambda _{2}z}+C_{3}e^{\lambda
_{3}z}+C_{4}e^{\lambda _{4}z}.
\end{equation}%
The perturbed ion velocity and density from the full system (\ref%
{full_system}) are found
\begin{align}
4\pi en_{i}& =\frac{\phi }{d_{e}^{2}}-\phi ^{\prime \prime }, \\
4\pi en_{0}v_{i}& =\frac{v_{0}}{d_{e}^{2}}\phi +\frac{c_{s}^{2}-v_{0}^{2}}{%
i\omega d_{e}^{2}}\phi ^{\prime }-v_{0}\phi ^{\prime \prime }+\frac{v_{0}^{2}%
}{i\omega }\phi ^{\prime \prime \prime }.
\end{align}%
The dispersion equation is obtained as a condition for the existence of a
nontrivial solution for $C_{1}$,$C_{2}$,$C_{3}$,$C_{4}$ in the linear system
of equations (\ref{boundary_conditions}) 
\begin{equation}
D=det%
\begin{pmatrix}
1 & 1 & 1 & 1 \\ 
e^{\lambda _{1}L} & e^{\lambda _{2}L} & e^{\lambda _{3}L} & e^{\lambda _{4}L}
\\ 
\lambda _{1}^{2} & \lambda _{2}^{2} & \lambda _{3}^{2} & \lambda _{4}^{2} \\ 
\mu _{1} & \mu _{2} & \mu _{3} & \mu _{4}%
\end{pmatrix}%
=0,  \label{DISP_matrix}
\end{equation}%
where 
\begin{equation}
\mu _{k}=\left( \frac{c_{s}^{2}}{v_{0}^{2}}-1\right) \lambda
_{k}+d_{e}^{2}\lambda _{k}^{3}.
\end{equation}

\begin{figure}[ht]
\includegraphics[width =140mm]{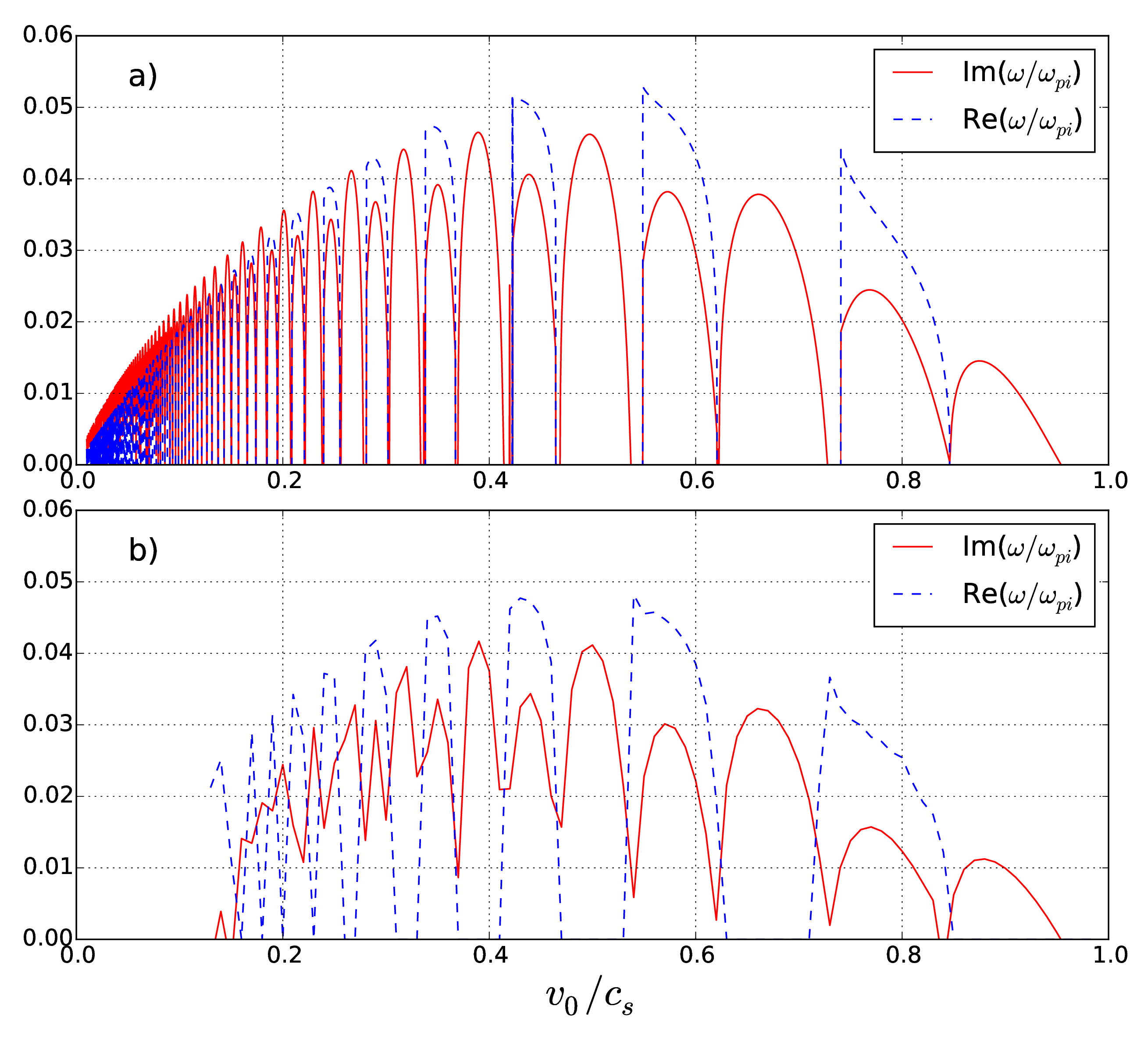}
\caption{The alternating zones of aperiodic ($\Re(\protect\omega) = 0$) and
oscillatory ($\Re(\protect\omega) \neq 0$) instabilities; a) - the solution
of the analytical dispersion equation (\protect\ref{DISP_matrix}), b) -
results of numerical simulations. }
\label{fig:week}
\end{figure}

The dispersion equation (\ref{DISP_matrix}) is difficult to solve
analytically as there are numerous solutions on the whole complex plane.
However we are interested only in those which have the largest imaginary
part, since these unstable modes will dominate. The numerical solution of
Eq. (\ref{DISP_matrix}) for the long system, with the length larger than the
Debye length, $L=10d_{e}$, is shown in Fig. \ref{fig:week}a. The mode
frequency is consistent with estimate (\ref{no_flow_aprox}).

For a fixed system length $L$, the instability growth rate depends on the
dimensionless ion flow velocity $v_{0}/c_{s}$. The unstable regions are
alternating with oscillatory ($\Re (\omega )\neq 0$) and aperiodic ($\Re
(\omega )=0$) zones. The boundaries of the zones could be found analytically
using the fact that at the boundary the wave frequency is zero. Expanding (%
\ref{DISP_matrix}) in Taylor series
\begin{equation}
D(\omega )=D(0)+\frac{\partial D(0)}{\partial \omega }\omega +O(\omega
^{2})=0,
\end{equation}%
and using that $D(0)\equiv 0$, $\frac{\partial D(0)}{\partial \omega }=0$
one finds 
\begin{equation}
\frac{v_{0}^2}{c_{s}^2}=\frac{1}{1+\pi ^{2}n^{2}\frac{d_{e}^{2}}{L^{2}}} 
\mbox{
where }n=1,2,3....  \label{zeros_wd}
\end{equation}%
solutions to this equation (\ref{zeros_wd}) correspond to zones boundaries
in Fig. \ref{fig:week}.

\subsection{Strong dispersion case}

In the short wavelength limit ($kd_{e}\gg 1\mbox{ or }d_{e}\gg L$), the
dispersion modifies the solution. In this limit, the ion sound modes are
reduced to the oscillations with the frequency of the order of $\omega \sim
\omega _{pi}$. 
In this case, the reciprocal of the Debye length ($1/d_{e}$) is considered
as a small parameter. Then the roots of the Eq. (\ref{eq_lambda}) are 
\begin{equation}
\lambda _{1,2}=0\mbox{ and }\lambda _{3,4}=i\frac{\omega \pm \omega _{pi}}{%
v_{0}},
\end{equation}%
and the general solution
\begin{equation}
\phi (z)=C_{1}\exp \left( i\dfrac{\omega +\omega _{pi}}{v_{0}}z\right)
+C_{2}\exp \left( i\dfrac{\omega -\omega _{pi}}{v_{0}}z\right) +C_{3}z+C_{4}.
\label{pir_sol_phi}
\end{equation}

This situation becomes mathematically equivalent to the Pierce instability.
Imposing boundary conditions (\ref{boundary_conditions}), one obtains an
homogeneous linear system, which has nontrivial solutions when the following
dispersion equation is satisfied
\begin{equation}
2\xi \alpha (1-e^{i\xi }cos\alpha )+i(\xi ^{2}+\alpha ^{2})sin\alpha e^{i\xi
}+i\dfrac{\xi ^{2}}{\alpha }(\xi ^{2}-\alpha ^{2})=0,  \label{disp_pierce}
\end{equation}%
where $\xi = L\omega /v_0$ and $\alpha = L\omega_p/ v_0$. \newline
It was shown \cite{Pierce,Mikh1}, that the dispersion equation (\ref%
{disp_pierce}) has following stability properties
\begin{subequations}
\label{p_zones}
\begin{align}
& \alpha <\pi & & \mbox{- has stable solution}, \\
& (2N-1)\pi <\alpha <2N\pi & & \mbox{- has aperiodic instability}, \\
& 2N\pi <\alpha <(2N-1)\pi & & \mbox{- has oscillatory instability},
\end{align}%
where $N=1,2,3....$, with a maximum growth rate $\gamma \sim v_0/L$.

There are many roots of the dispersion equation (\ref{disp_pierce}) on whole
complex plane; as before, we choose only roots which have the largest
imaginary part. The solutions which meets these criteria are shown in Fig. %
\ref{fig:pirce}. The alternating aperiodic and oscillatory instability zones
exist similar to the weak dispersion case. Fig. \ref{fig:week} also shows
the results of direct initial value simulations described in the next
section.

\begin{figure}[ht]
\centering
\includegraphics[width = 140mm]{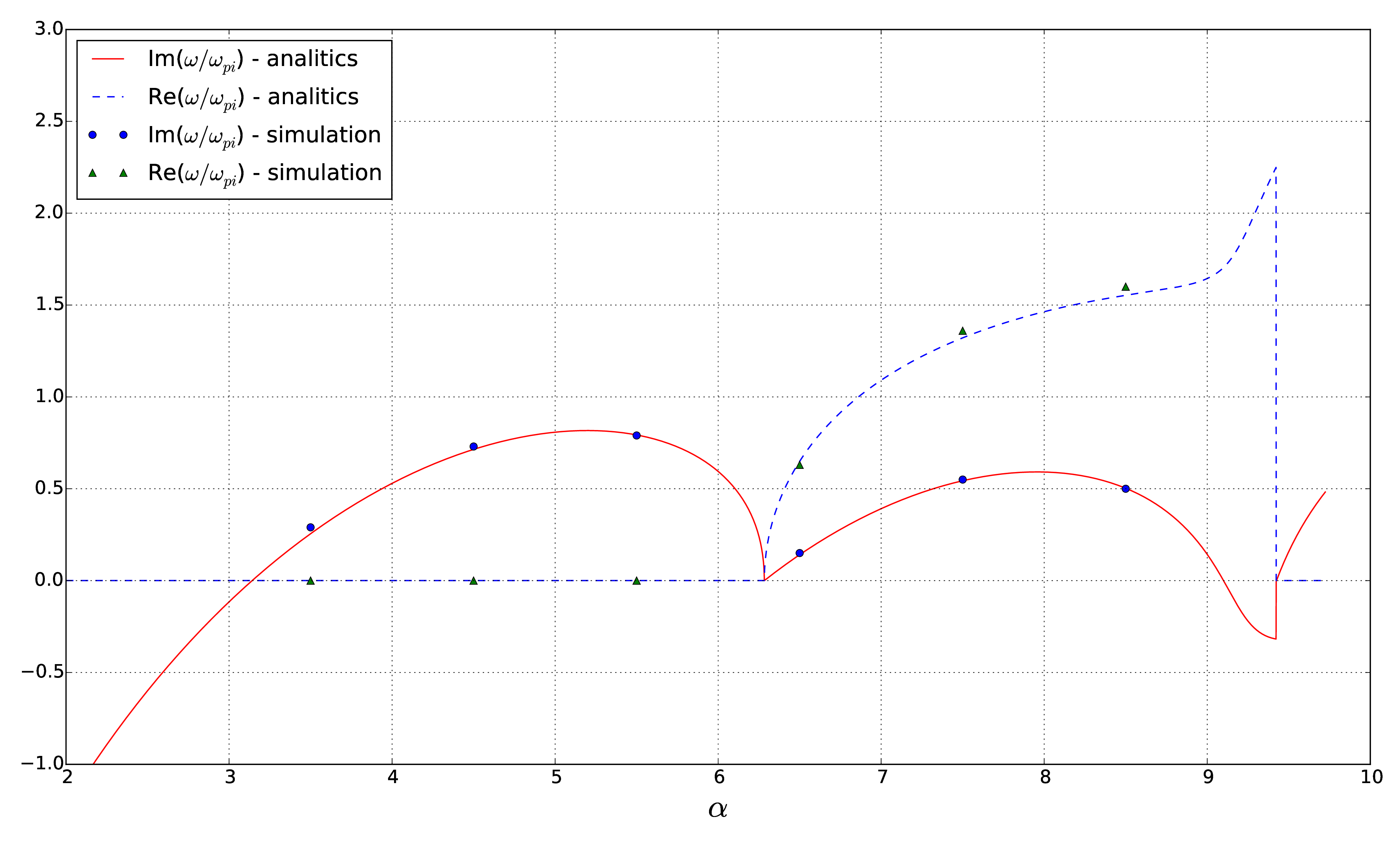}
\caption{ The oscillatory ($\Re(\protect\omega) \neq 0$) and aperiodic ($\Re(\protect\omega) = 0$) instabilities in strong dispersion case. The
analytical solution of (\protect\ref{disp_pierce}) and numerical simulations for $L = 0.1d_e$. }
\label{fig:pirce}
\end{figure}

\section{Numerical solution}

To confirm our analytical results we solve the system (\ref{full_system})
numerically. These equations have different structure and we employ the
following strategy. The first two equations of (\ref{full_system}) are
considered as an explicit initial value problem (IVP), and the the third and
fourth equations of (\ref{full_system}) are solved as a boundary value
problem (BVP). These subsystems are solved numerically in time and to obtain
the time dependent evolution of IVP and BVP they are solved iteratively. The
Poisson equation in BVP is solved at the beginning of each time step. The
BVP system uses the given ion density profile (either from the initial
condition or from previous time step) to produce the electrostatic potential
profile. The known potential distribution allows us to solve IVP in time. As
the final step, we update, the ion density and velocity profiles obtained
from IVP.

Common ways to solve a BVP\cite{Numerical_Recipes} are a family of shooting
methods and finite difference schemes. We use shooting methods due to their
simplicity. We have selected multiple shooting method (MSM)\cite{Ascher_BVP}
because it is easy to parallelize, it has no disadvantages of simple
shooting methods (e.g., limitations on a system length).

Our IVP is a system of hyperbolic partial differential equations (PDEs), which can be expressed in a
conservative form, because of the nature of the continuity and Euler
equations, which are conservative. This suggests to treat our system with a
class of finite volume methods \cite{FVM}. The simplest finite volume method
is an upwind scheme, however we cannot use this scheme for all situation
because our physical model contains the waves propagating in opposite
directions that will make the upwind unconditionally unstable. Therefore, we
have resorted to Harten, Lax, Van Leer (HLL) \cite{HLL} belonging to the
Godunov family methods. Such schemes can be characterized by the solution of
Riemann problem on computational cells. There are two types of Godunov methods: approximate and exact Riemann solvers. We used one of the kind of approximate Riemann solves
- the HLL method.

For convenience all further results will be expressed in dimensionless
units
\end{subequations}
\begin{equation}
\frac{n}{n_{0}}\rightarrow n\mbox{,\hspace{0.5cm}}\frac{z}{d_{e}}\rightarrow
z\mbox{,\hspace{0.5cm}}\frac{e\phi }{T_{e}}\rightarrow \phi %
\mbox{,\hspace{0.5cm}}t\omega _{pi}\rightarrow t\mbox{,\hspace{0.5cm}}\frac{v%
}{c_{s}}\rightarrow v\mbox{,\hspace{0.5cm}}\frac{L}{d_{e}}\rightarrow L%
\mbox{,\hspace{0.5cm}}\frac{v_{0}}{c_{s}}\rightarrow v_{0}.
\end{equation}%
The results of numerical simulations are compared with analytical results
for week and strong dispersion cases. We start our simulations with initial
conditions of a uniformly distributed random noise and observe the
evolution of the following quantities
\begin{equation}
N^{2}=\int_{0}^{L}n^{2}(z)dz,\mbox{ }\Phi ^{2}=\int_{0}^{L}\phi ^{2}(z)dz,%
\mbox{ }V^{2}=\int_{0}^{L}v^{2}(z)dz.
\end{equation}%
Depending on the value of input parameters ($L,v_{0}$) damped (stable) or
growing (unstable) solutions were observed. Unstable solution were fitted to
the following curves 
\begin{equation}
N^{2},V^{2},\Phi ^{2}\sim \cos (2\Re (\omega )t+\theta )e^{2\gamma t},
\end{equation}%
to determine the real frequency and growth rate.

\begin{figure}[ht]
\centering
\includegraphics[width = 140mm]{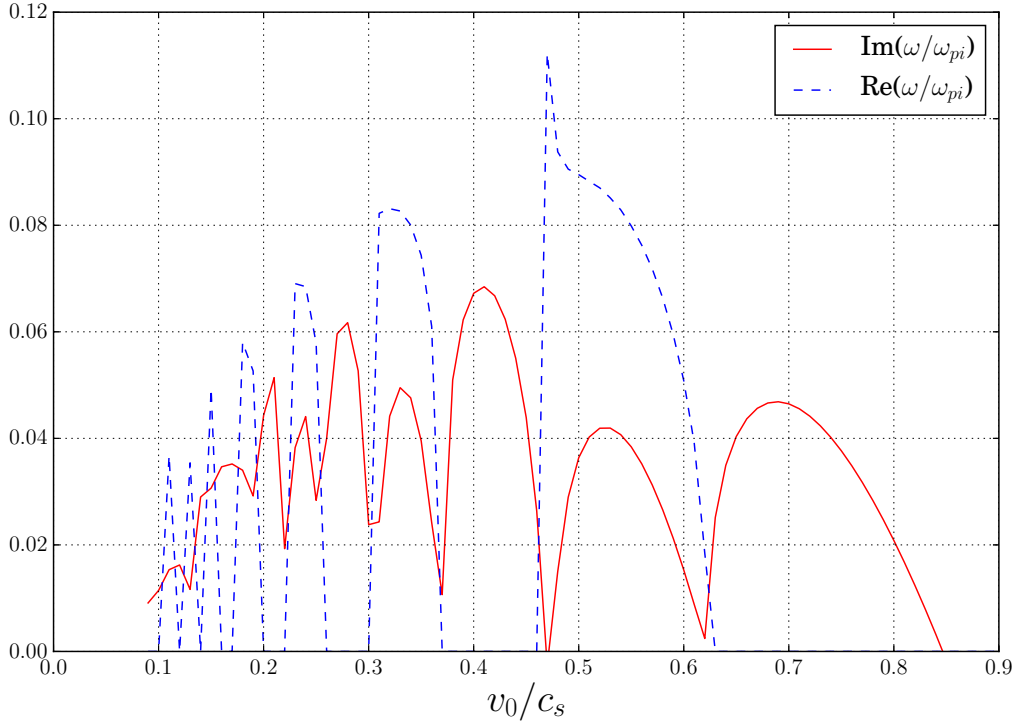}
\caption{ Alternating oscillatory ($\Re(\protect\omega) \neq 0$) and
aperiodic ($\Re(\protect\omega) = 0$) instabilities zones in the
intermediate system length $L=5$; numerical simulations results.}
\label{fig:L5}
\end{figure}

When the length of the system exceeds the Debye length ($L\sim 10d_{e}$),
the week dispersion results are recovered. Example of frequency and growth
rate dependence as a function of the ion flow velocity $v_{0}$ are shown in
Figs. \ref{fig:week}b and \ref{fig:L5}. These graphs are similar to the
analytical results shown in Fig. \ref{fig:week}a. In fact, the difference of
the analytical and numerical results are of the order of the magnitude of
the small parameter of the analytical theory ($d_{e}/L$). Due to the
increasing density of the instability zones, very high resolution is
required to recover the singular part ($v_{0} \rightarrow 0$) of the analytical
solution.

From the theory we know that instability will not occur in quasi-neutral
case. In other words, charge separation is crucial for the instability to
occur. Because in the long system charge separation is less prominent we can expect decreasing of instability
growth rate with system length. This is confirmed by simulations for $L=5$
(Fig. \ref{fig:L5}) and $L=10$ (Fig. \ref{fig:week}).

In the regime, when the length of the system is much smaller than the Debye
length ($L\sim 0.1d_{e}$) the difference between analytical solution of
strong dispersion approximation and numerical solution was less than few
percent. This comparison is shown in Fig. \ref{fig:pirce}.

\begin{figure}[ht]
\subfloat[zone \#1 $v_0=0.9$]{\includegraphics[width = 54mm]{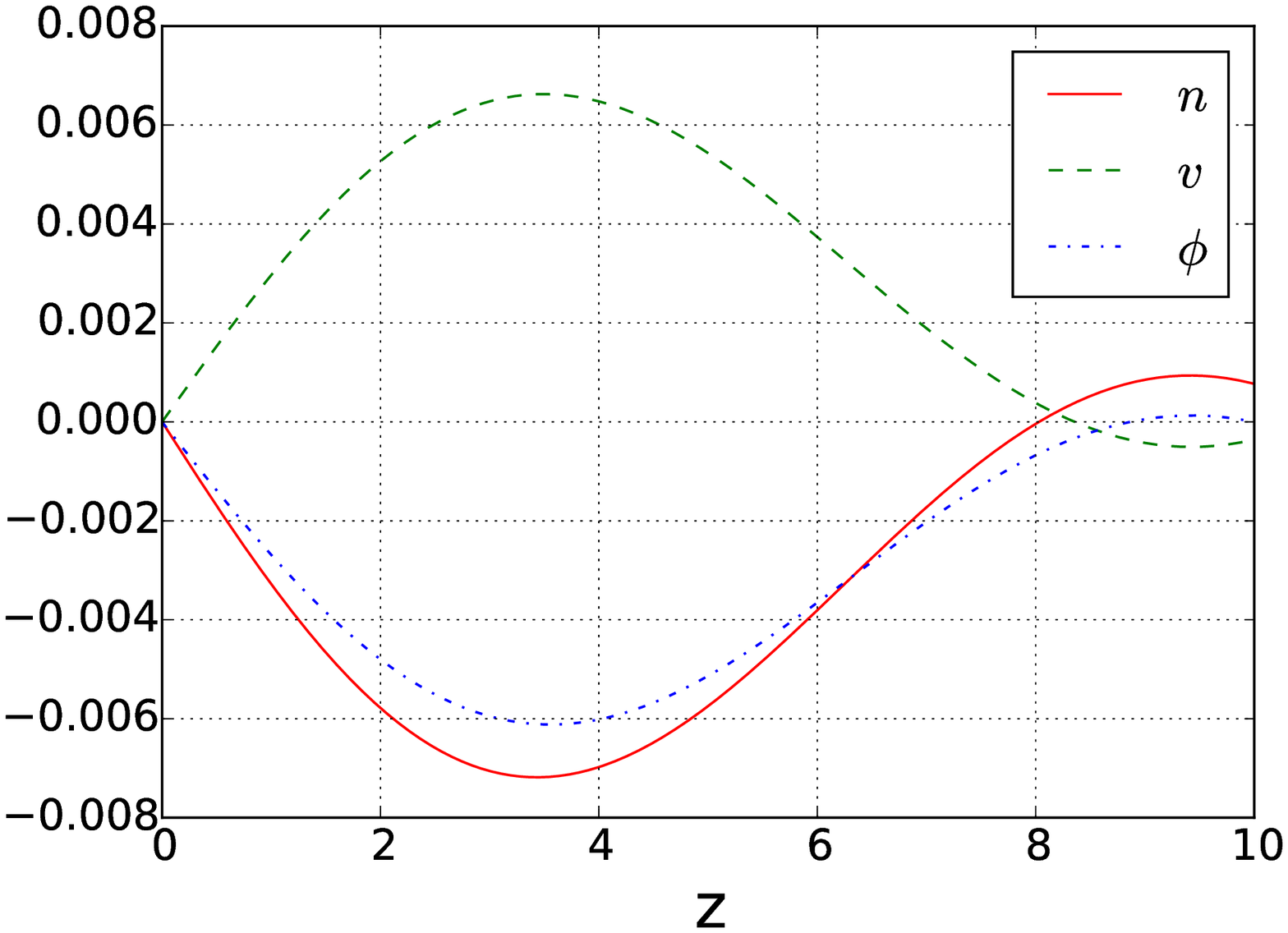}} %
\subfloat[zone \#2 $v_0=0.78$]{\includegraphics[width = 54mm]{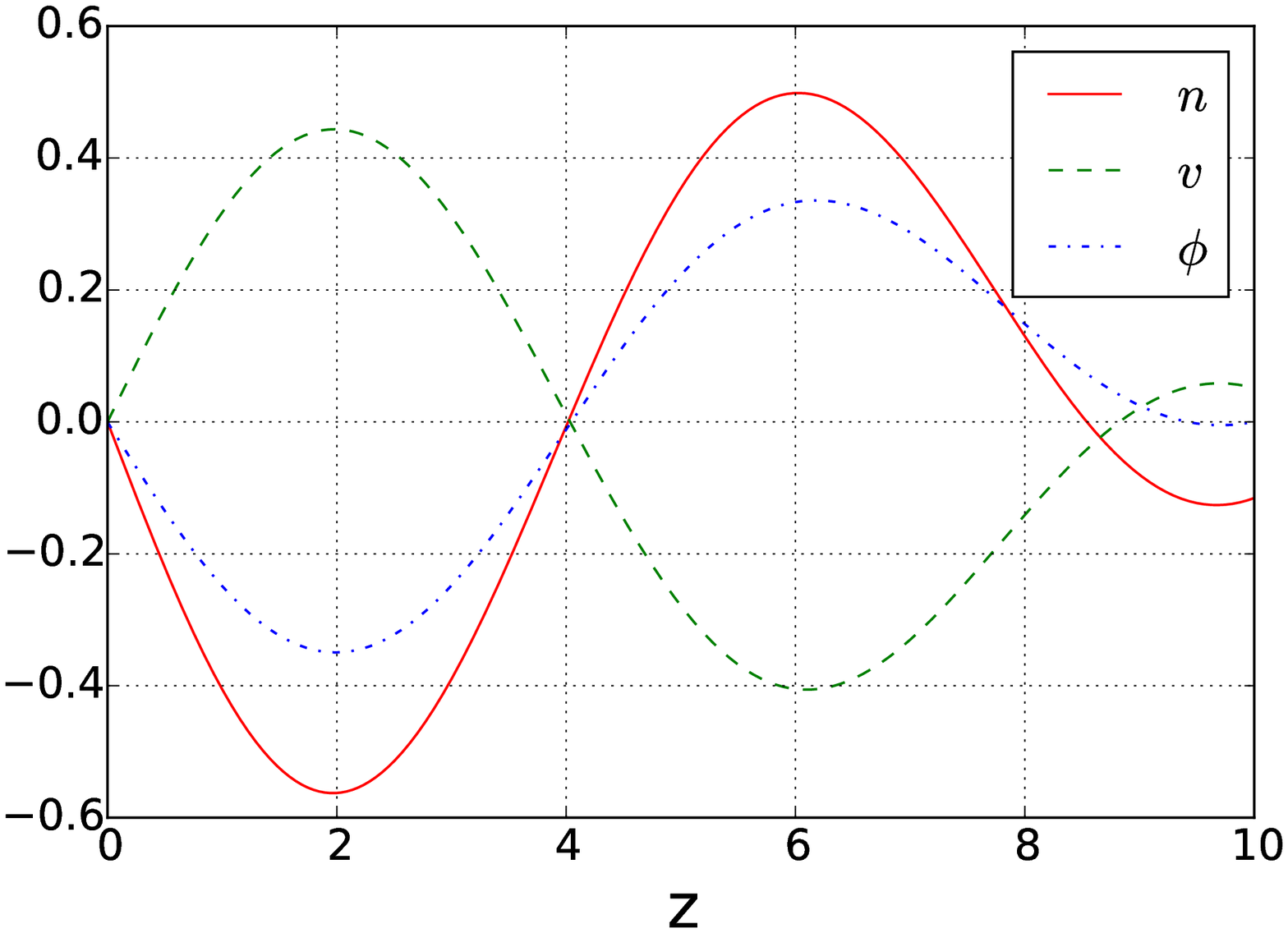}} %
\subfloat[zone \#3 $v_0=0.65$]{\includegraphics[width = 54mm]{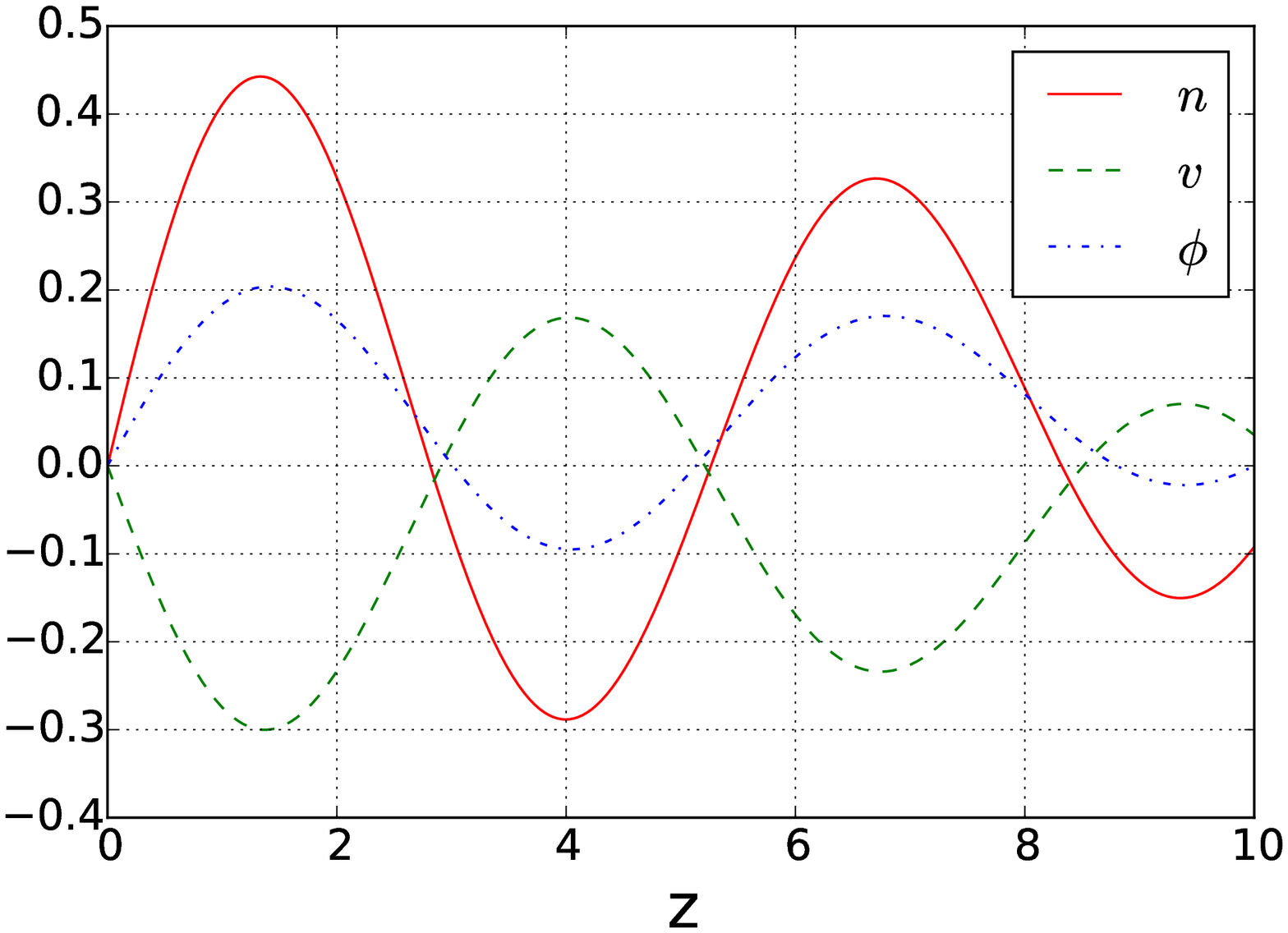}}
\caption{Unstable spatial eigenfunctions of density, velocity and
electrostatic potential for $L=10$, for different instability zones from
Fig. \protect\ref{fig:week}. Zone numbers in Fig. \protect\ref{fig:week} are
counted from the right, with the right outermost aperiodic zone as \#1. }
\label{fig:zones}
\end{figure}

The number of zeros of unstable spatial eigenfunctions of density, velocity
and electrostatic potential correlates with the zone number (\ref{p_zones})
and is defined by the value of the $\alpha $ parameter. In more
general case, the stability of the system is governed by two parameters ($%
v_{0},L$). However in general case
the number of zeros correlates with a number of zone as well, examples of
eigenfunctions are shown on Fig. \ref{fig:zones}.

\begin{figure}[ht]
\centering
\includegraphics[width=140mm]{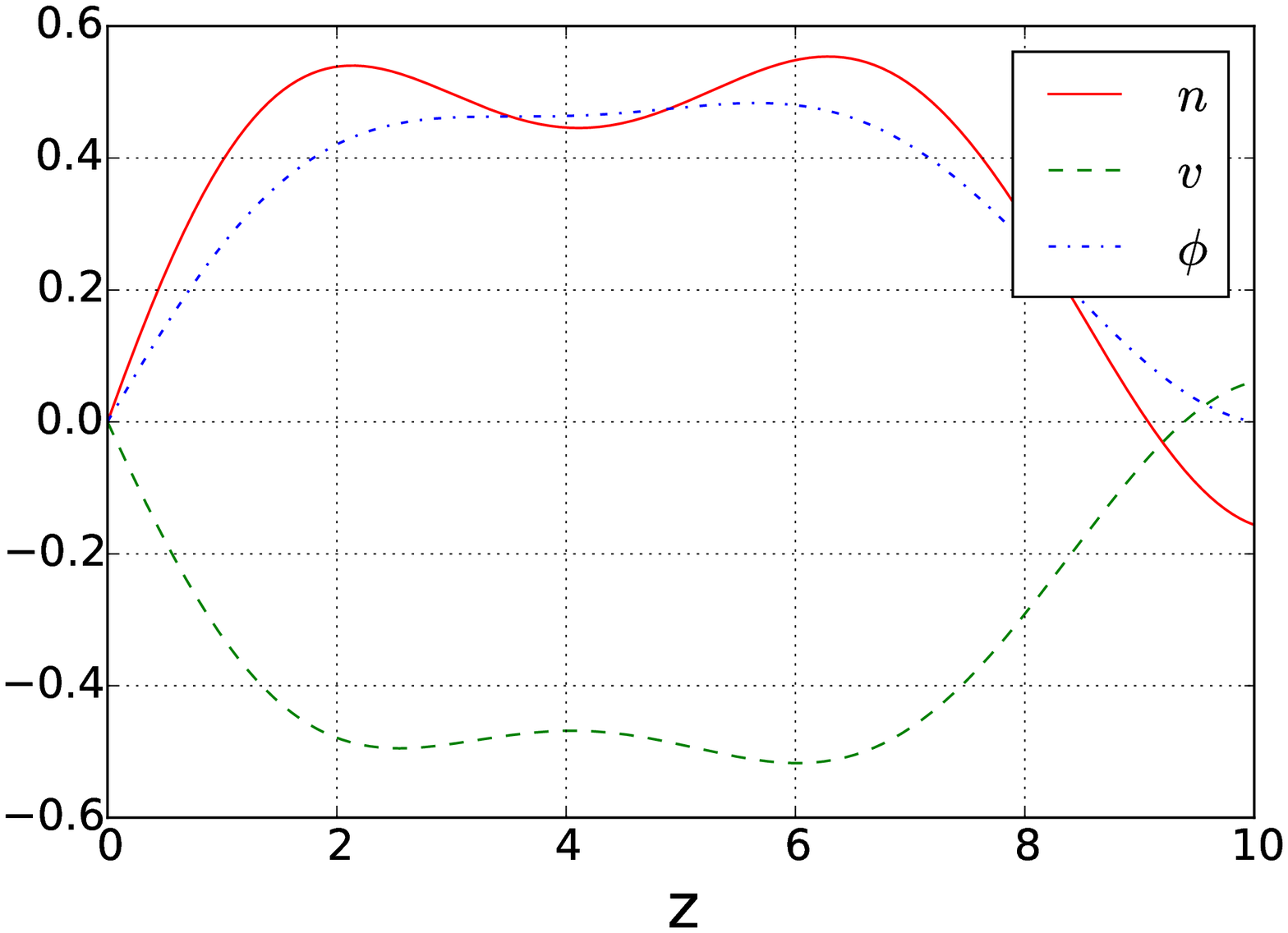}
\caption{Unstable eigenfunctions of density, velocity and electrostatic
potential in the oscillatory zone \#2 $v_0=0.78$, $L=10$, at a later time. }
\label{fig:zero_br}
\end{figure}

In aperiodic zones (where real part of frequency is zero) the number of
nodes does not change during the time evolution. In oscillatory zones, some
nodes disappear at later times as shown in Fig. \ref{fig:zero_br}.

\begin{figure}[ht]
\subfloat[t = 0]{\includegraphics[width = 70mm]{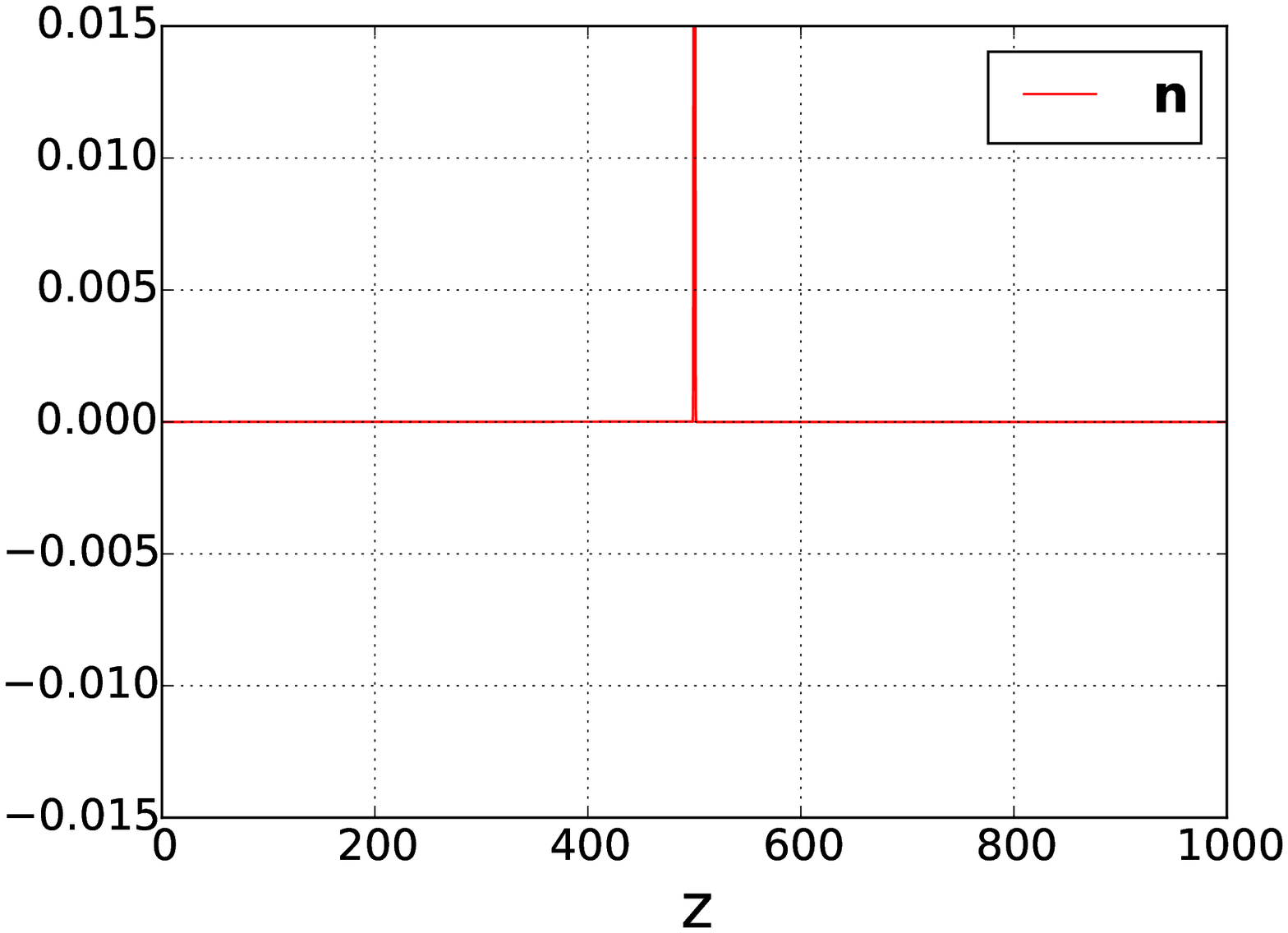} \label{fig:ev:wd:1}}
\subfloat[t = 200]{\includegraphics[width = 70mm]{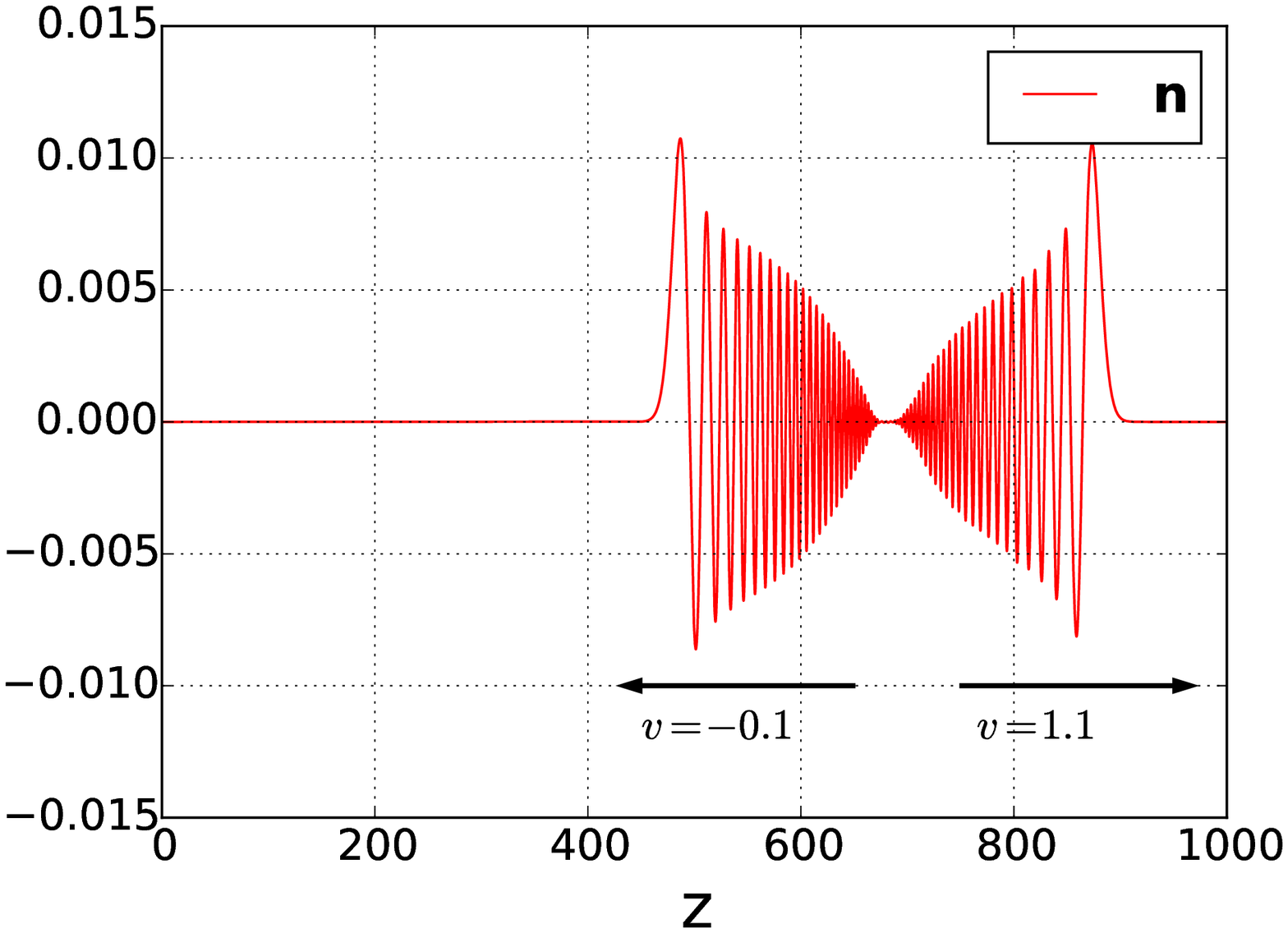}
\label{fig:ev:wd:2}} \newline
\subfloat[t = 300]{\includegraphics[width = 70mm]{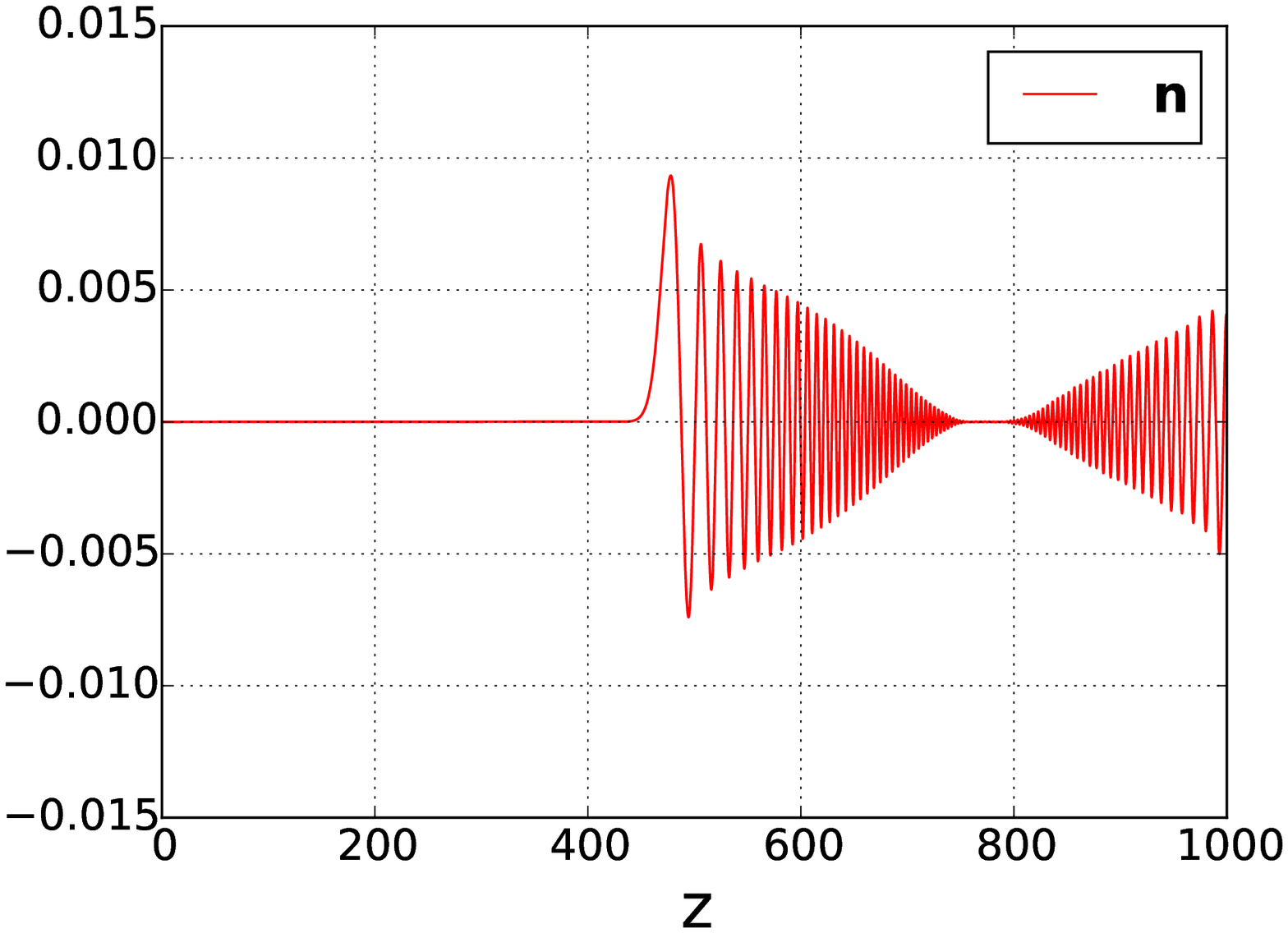}
\label{fig:ev:wd:3}} 
\subfloat[t = 600]{\includegraphics[width =
70mm]{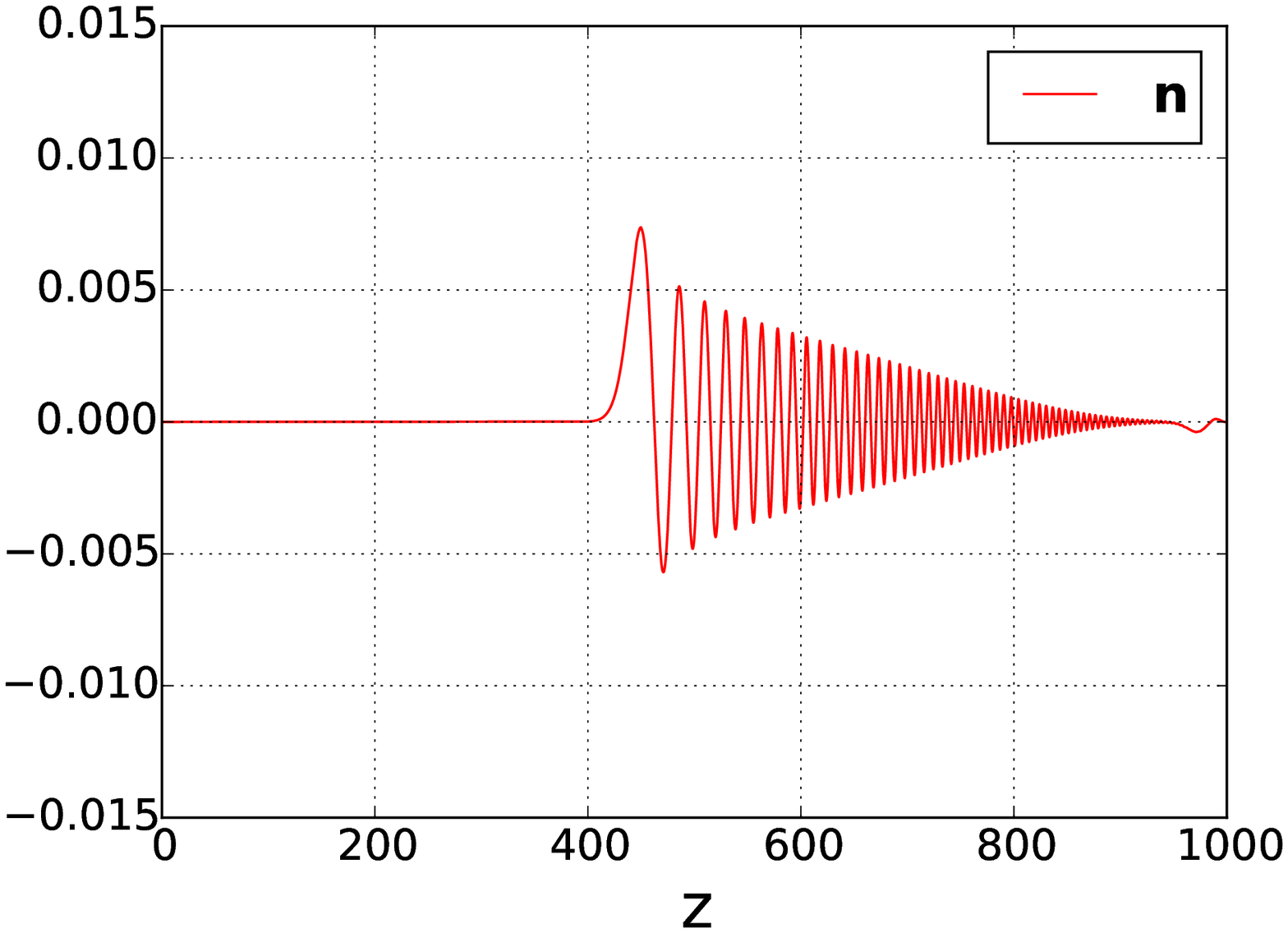} \label{fig:ev:wd:4}} \newline
\subfloat[t = 4400]{\includegraphics[width = 70mm]{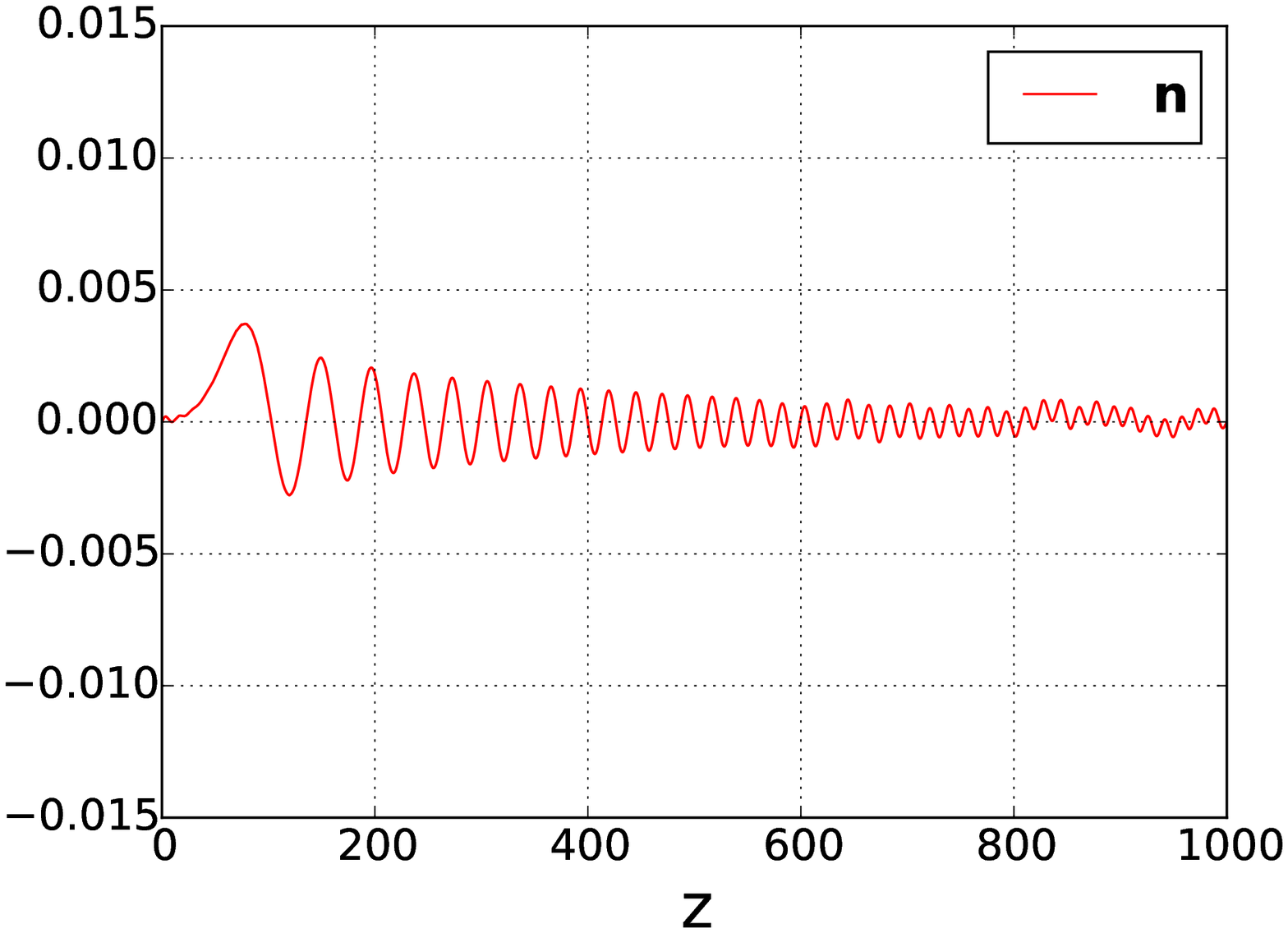}
\label{fig:ev:wd:5}} 
\subfloat[t = 5100]{\includegraphics[width =
70mm]{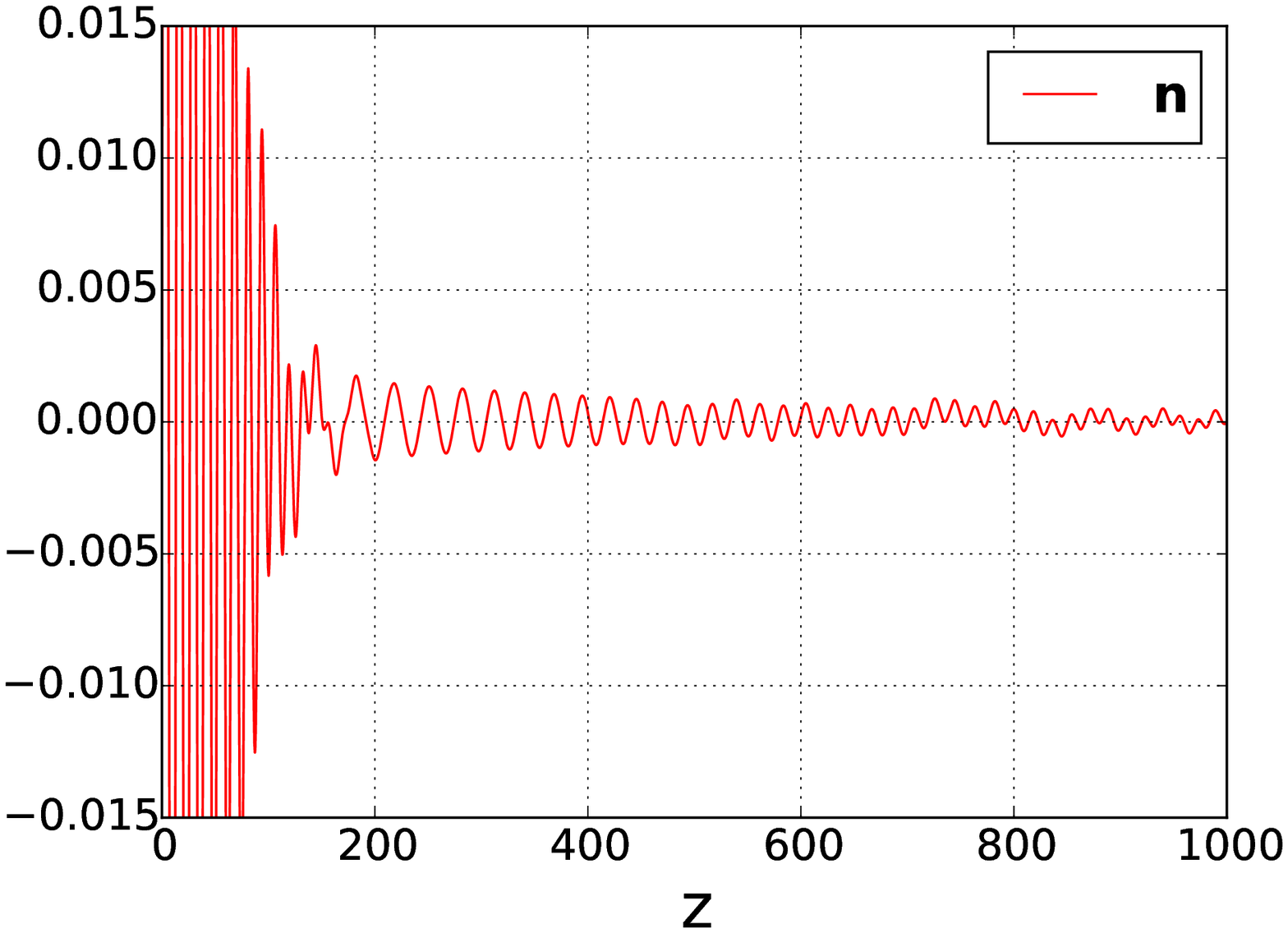} \label{fig:ev:wd:6}}
\caption{Evolution of the initial Gaussian pulse in the weak dispersion
case: (a) - initial condition; (b) - initial perturbation splits into two
traveling wave packets, the one traveling to the right with $v_0+c_s=1.9$
and the one traveling to the left with $v_0-c_s=-0.1$; (c) - the right wave
packet is passing through the right wall barely reflecting; (d) - the
beginning of the reflection of the left wave packet from the wall and
forming of the unstable eigen-function.}
\label{fig:wd_wave_travel}
\end{figure}

In weakly dispersive case ($kd_{e}\ll 1$), the addition of the Doppler shift
due to the ion flow velocity results in the main order modification for the
propagating modes velocities 
\begin{equation}
v_{1,2}=v_{0}\pm c_{s},  \label{ph_vel_wd}
\end{equation}%
which correspond the one pair of the of the roots of Eq. (\ref{eq_lambda}).
Two other roots describe the slow dispersion effects. We have chosen very
long system ($L=1000d_{e}$) so the dispersion is weak and two wave packets
are well separated. Gaussian function localized in the middle of the system
was chosen as an initial condition, Fig. \ref{fig:ev:wd:1}. Fig. \ref%
{fig:ev:wd:2} shows that the Gaussian peak separated into two wave packets
moving in opposite directions with velocities $v_{1,2}$ from Eq. (\ref%
{ph_vel_wd}). The right wave packet meets the wall at the right and passes
through the wall with almost no reflection, as shown in Figs. \ref%
{fig:ev:wd:3} and \ref{fig:ev:wd:4}. Instability occurs when the slow wave
packet meets the left wall (with Dirichlet boundary conditions for all
variables) and is reflected, Fig. \ref{fig:ev:wd:5}. At a later time, the
reflected wave and dispersion tail overlap forming an unstable
eigenfunction, Fig \ref{fig:ev:wd:5}.

\begin{figure}[ht]
\subfloat[t = 0]{\includegraphics[width = 80mm]{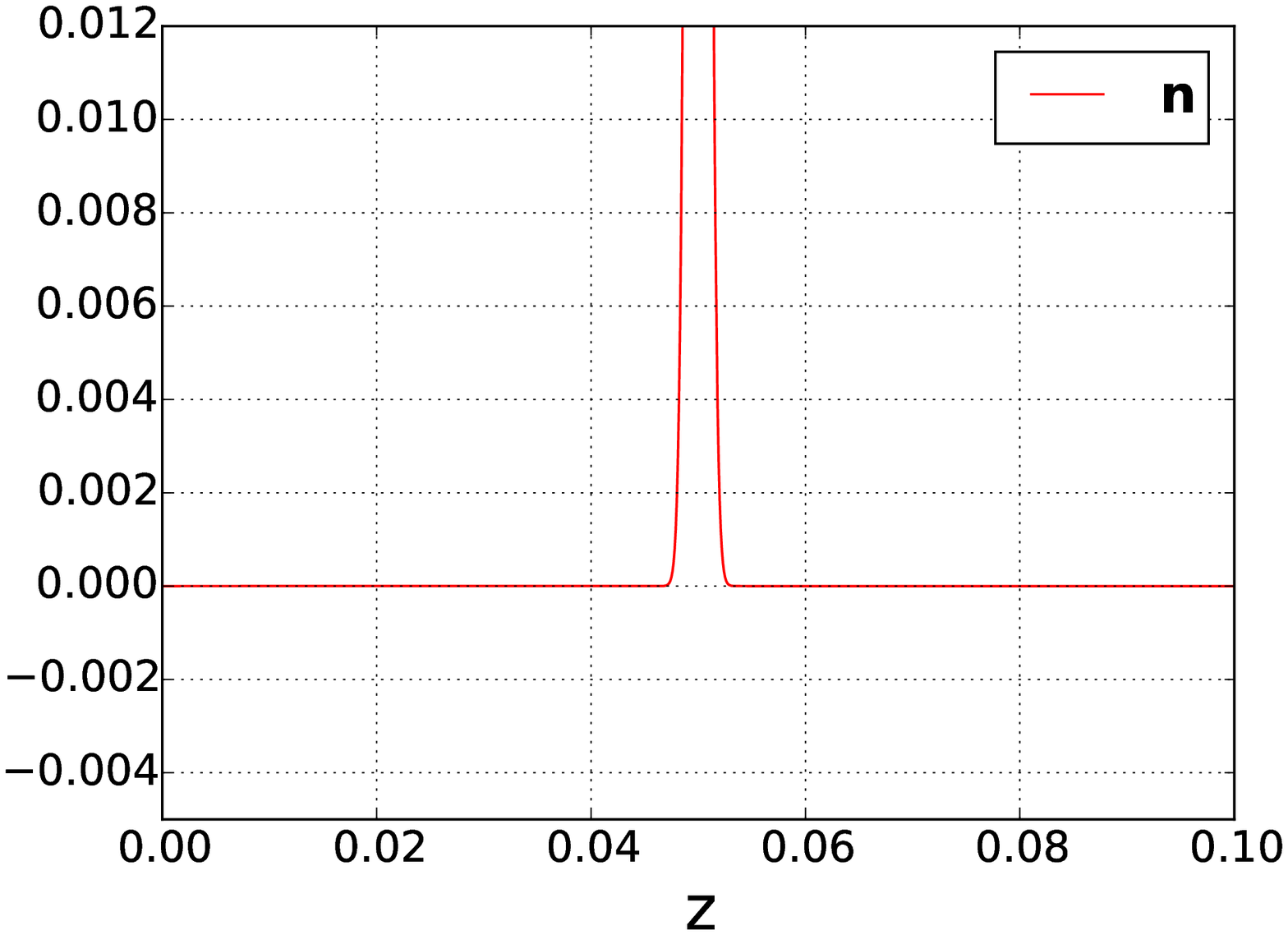} \label{fig:ev:sd:1}}
\subfloat[t = 100]{\includegraphics[width = 80mm]{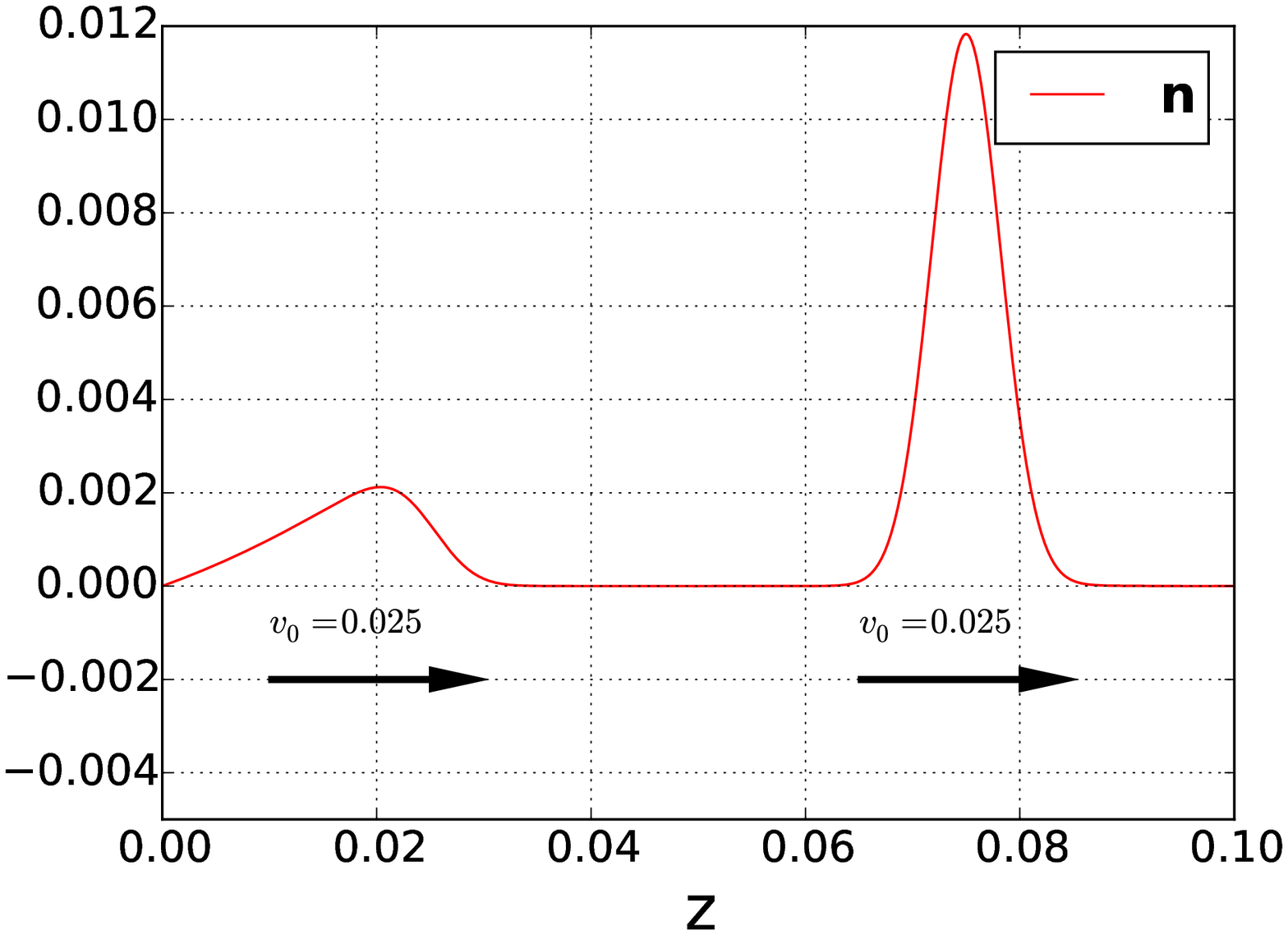}
\label{fig:ev:sd:2}} \newline
\subfloat[t = 300]{\includegraphics[width = 80mm]{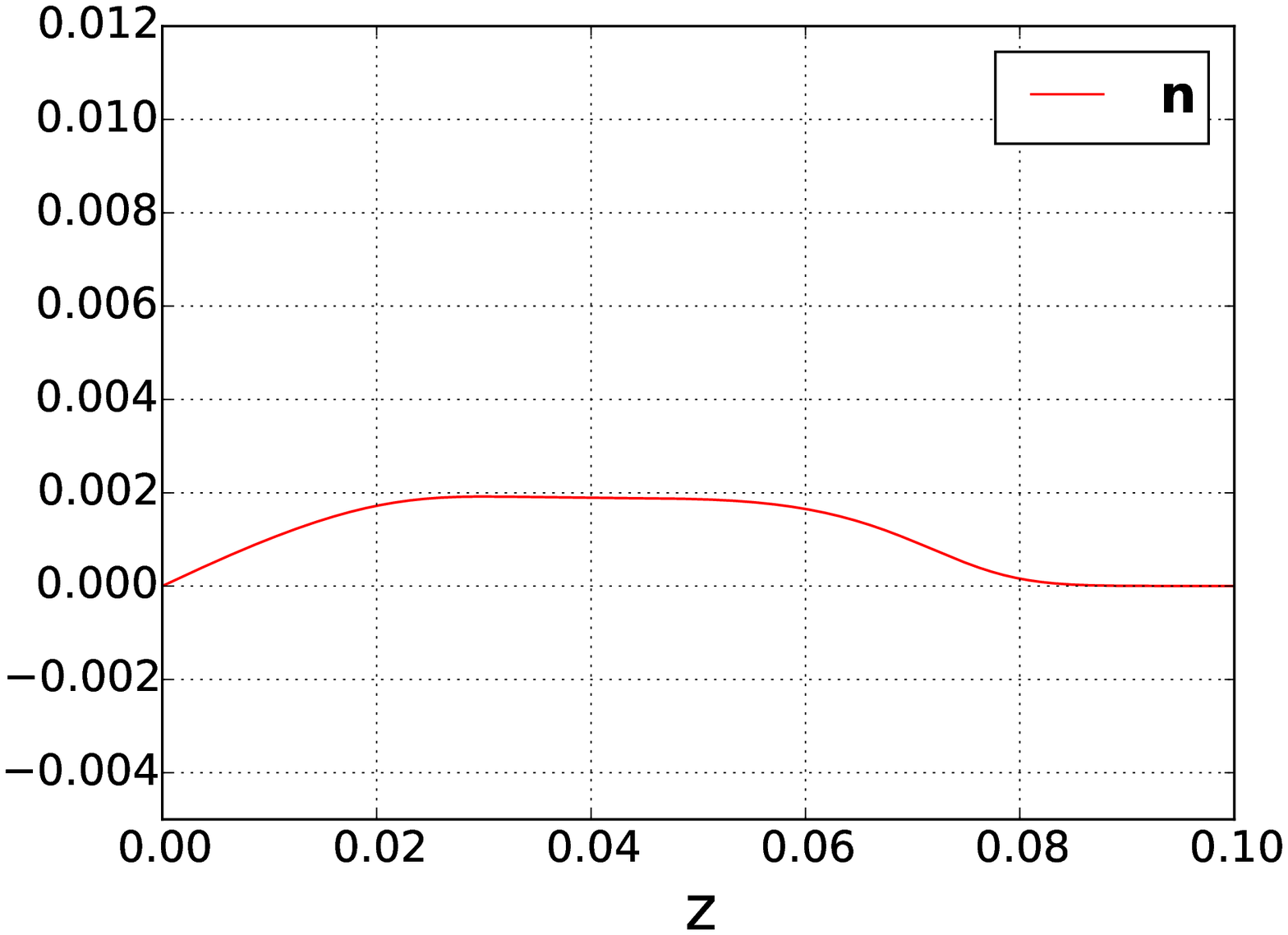}
\label{fig:ev:sd:3}} 
\subfloat[t = 900]{\includegraphics[width =
80mm]{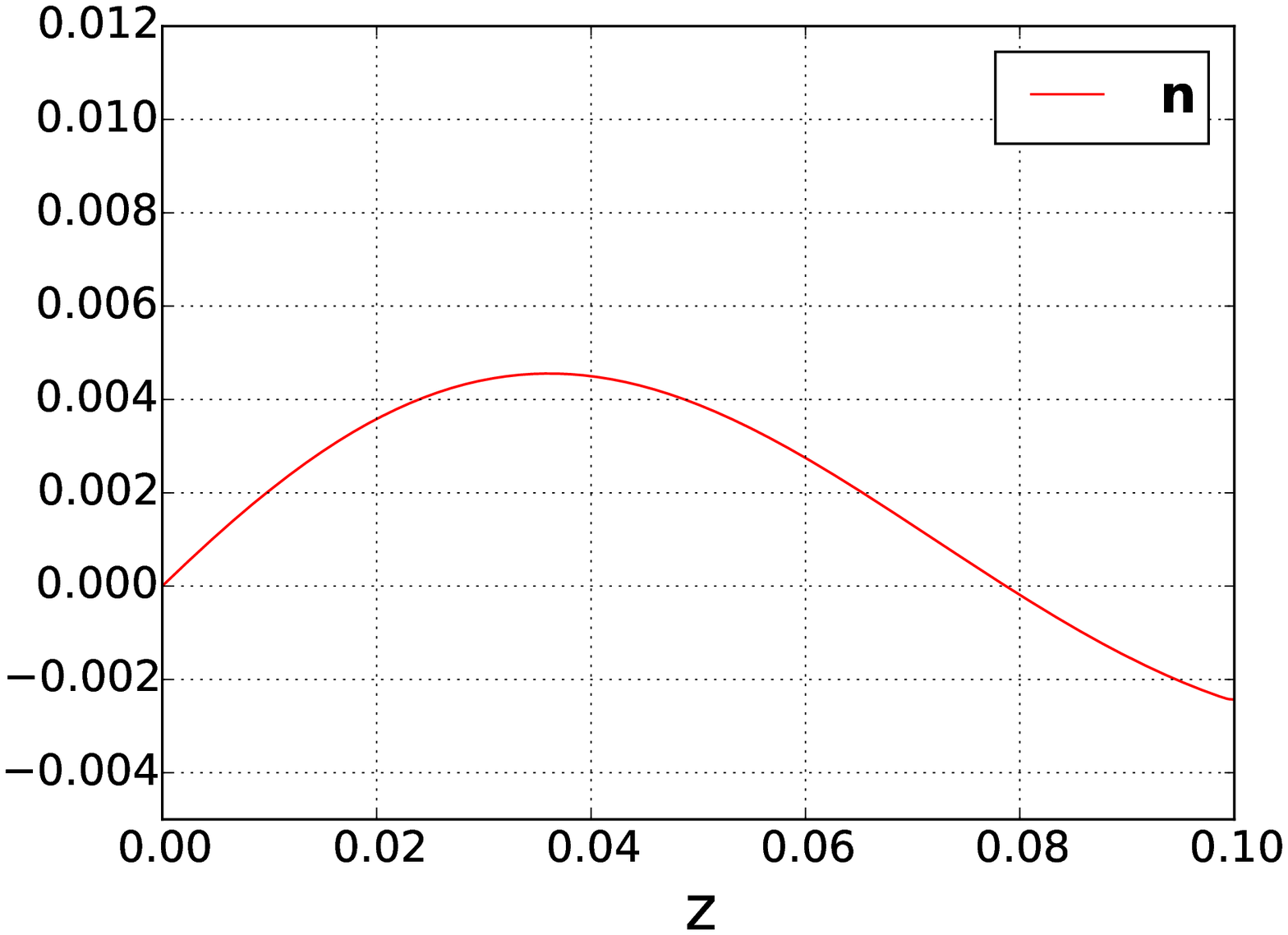} \label{fig:ev:sd:4}}
\caption{Dynamics in the strong dispersion case: (a) - initial state; (b) -
the initial Gauss pulse travels with velocity $v_0 = 0.025$ to the right,
another pulse start to grow and travels to the right with the same velocity;
(c) - the initial pulse approaches the left boundary and the unstable
eigen-function forms. }
\label{fig:sd_wave_travel}
\end{figure}

In strong dispersion case the equation (\ref{no_flow_de}) implies that
oscillations with the ion plasma frequency will occur. The short system was
chosen ($L=0.1d_{e}$) to demonstrate this regime. Initial condition was
chosen in the form of the Gaussian function localized in the middle. The
evolution is shown in Figs. \ref{fig:sd_wave_travel}. First frame is an
initial Gaussian peak which travels with velocity of the ion flow ($v_{0}$);
at the same time another peak arises from the left border and starts to
travel with same velocity. Note that in case of strong dispersion, the ions
sound phase velocity is much reduced$,\omega /k<c_{s}$. When the initial
Gaussian peak meets the right boundary (which has no boundary conditions
except the one for electrostatic potential) it passes through, while another
peak starts to transform to unstable eigenfunction at the left boundary.

\section{Conclusion}

We have investigated the ion acoustic instability induced by the ion flow in
a finite length system; the situation which is relevant to various plasma
devices such as electric propulsion and emissive probe diagnostics. It was
shown that the length of the system measured in units of the Debye length
and ion flow velocity measured in units of the ion acoustic velocity are
important parameters which control the instability. \newline
For long systems ($d_{e}\ll L$) the analytical dispersion equation was
obtained describing the aperiodic and oscillatory
instability zones. The boundaries of the instabilities are defined by the
condition (\ref{zeros_wd}). The instability criteria could also be written
in the form 
\begin{equation}
\frac{1}{1+\pi ^{2}\frac{d_{e}^{2}}{L^{2}}}>\frac{v_{0}^{2}}{c_{s}^{2}}.
\label{my_cond}
\end{equation}%
For short systems ($d_{e}\gg L$) the dispersion equation was obtained in the
form equivalent to the Pierce dispersion equation. In this case, the
following instability criteria has been obtained 
\begin{equation}
L \omega_{pi}/ \pi > v_{0}.
\end{equation}
Analytical theory was confirmed by the results of direct initial value
numerical simulations. We have investigated the structure of the
eigenfunctions in the unstable zones. It is shown that the order of the
instability zone correlates with a number of nodes in the corresponding \
eigenfunction. Our numerical simulations show that the instability occurs as
a result of the mode coupling mediated by the boundaries. \newline

The instability mechanism in a finite length system is different from the
kinetic ion sound instability \cite{akhiezer} in infinite plasmas. The
dispersion equation for the latter can be written in the form 
\begin{equation}
1+\frac{\omega _{pi}^{2}}{k^{2}c_{s}^{2}}-\frac{\omega _{pi}^{2}}{\omega
-kv_{0}}+i\sqrt{\frac{\pi }{2}\frac{m_{e}}{m_{i}}}\frac{\omega
_{pi}^{2}\omega }{k^{3}c_{s}^{3}}=0.
\end{equation}%
Treating $\epsilon =\sqrt{\frac{\pi }{2}\frac{m_{e}}{m_{i}}}$ as a small
parameter, one obtains the growth rate
\begin{equation}
\gamma =\frac{\epsilon kc_{s}}{2(1+k^{2}d_{e}^{2})^{2}}\left( -1\pm \frac{%
v_{0}}{c_{s}}\sqrt{1+k^{2}d_{e}^{2}}\right) .
\end{equation}%
The instability condition has a form 
\begin{equation}
\frac{1}{1+k^{2}d_{e}^{2}}<\frac{v_{0}^{2}}{c_{s}^{2}},  \label{kin_cr}
\end{equation}%
which is complementary to the condition (\ref{my_cond}).

The excitation of large scale perturbation and soliton formation was
observed in a number of experiments \cite{soliton_lonngren,soliton_hirose}.
Similar structures may be excited by ion flow due to the mechanism
identified in our paper which is operative in systems of a finite length and
in situations when the ion flow velocity is below the ion acoustic speed.
The excitation of ion sound waves in a finite length system was observed in
numerical particle-in-cell simulations with emissive walls \cite{Sydorenko1,Sydorenko2}.
The mechanism described in this paper can also be relevant to the
instabilities observed in double layer experiments \cite{DL1,DL3,DL4}. 

\bigskip 

\textbf{Acknowledgments}

This work was supported in part by NSERC\ of Canada and US Air Force Office
of Scientific Research.

\end{document}